\documentclass[aps,prc,preprint,groupedaddress,onecolumn,floatfix]{revtex4}
\pdfmapfile{-mpfonts.map}
\usepackage{amsmath}
\usepackage{amssymb}
\usepackage{amsfonts}
\usepackage{amscd}
\usepackage{bm}
\usepackage{bbold}
\usepackage{inter}
\usepackage{graphicx}
\usepackage{color}
\usepackage{empheq}

\def\beq{\begin{equation}}
\def\eeq{\end{equation}}
\begin{document}

\title{Perturbation theory of the continuous spectrum in the theory of nuclear reactions}

\author{Brady J. Martin  and
W.~N.~Polyzou  
}
\email{polyzou@uiowa.edu}
\thanks{This work supported by the U.S. Department of Energy,
  Office of Science, Grant \#DE-SC16457}
\affiliation{Department of Physics and Astronomy\\
  The University of Iowa\\
  Iowa City, IA 52242, USA}


\begin{abstract}
{\bf \,}
\bigskip
  
\noindent {\bf Background:} 
Nuclear reactions are complex, with a large number of possible
channels.  Understanding how different channels contribute to a given
reaction is investigated by perturbing the continuous spectrum.

\bigskip

\noindent{\bf Purpose:}
To develop tools to investigate reaction mechanisms by
identifying the contributions from each reaction channel.

\bigskip

\noindent{\bf Method:}
Cluster decomposition methods, along with the spectral theory of proper
subsystem problems, is used to identify the part of the nuclear Hamiltonian
responsible for scattering into each channel.

\bigskip

\noindent{\bf Results:}
The result is an expression of the nuclear Hamiltonian as a sum over
all scattering channels of channel Hamiltonians.  Each channel
Hamiltonian is constructed from solutions of proper subsystem
problems.  Retaining any subset of channel Hamiltonians results in a
truncated Hamiltonian where the scattering wave functions for the
retained channels differ from the wave functions of the full
Hamiltonian by $N$-body correlations.  The scattering operator for the
truncated Hamiltonian satisfies an optical theorem in the retained
channels.  Because different channel Hamiltonians do not commute, how
they interact determines their contribution to the full dynamics.

\end{abstract}

\maketitle

\section{Introduction}
\label{intro}

\noindent The exact solution of the many-body scattering problem is an
intractable numerical problem due to the large number of degrees of
freedom.  Even if it could be solved numerically, the solution does
not determine the role of different channels and how they interact in
a given reaction.  The ability to examine the response to turning
different channels on or off provides means to investigate the impact
of that channel to a complicated reaction.  This is useful for
determining which channels are most responsible for nuclear binding,
decays, or a given cross section.  Mathematically, this is a problem
in perturbation theory of the continuous spectrum.  This is not
trivial from a mathematical point of view because the continuous
spectrum is not stable with respect to small perturbations
\cite{Weyl}\cite{JvNeumannW}\cite{Kato}.  Formal scattering theory
provides tools to study the continuous spectrum.  The optical theorem
provides a means to isolate contributions from different channels.

In this work, the exact many-body Hamiltonian is expressed as a linear
combination of non-commuting self-adjoint operators associated with
each scattering channel.  These are referred to as channel
Hamiltonians.  Each channel Hamiltonian is constructed using only
proper subsystem solutions.  Time-dependent scattering theory is used
to show that each channel Hamiltonian has scattering solutions in that
channel.  Sums of subsets of channel Hamiltonians result in a
truncated Hamiltonian that satisfies an optical theorem in that subset
of channels.  The resulting scattering wave functions agree
with the exact scattering wave functions up to, but not including,
$N$-body correlations.  Because the channel Hamiltonians do not commute,
including additional channels in the sum affects the dynamics of each retained
channel, but the additional channels only impact the $N$-body
correlations in the channel wave function.

Since $N$-particle bound states are determined by the Hamiltonian, the
expression of the full Hamiltonian as a sum of channel
Hamiltonians provides a means to determine how each channel
Hamiltonian contributes to the binding.  It can also happen that a
subset of scattering channels can support a bound state, which may become
unstable as more channels are added.

This problem was formally solved in \cite{Polyzou:1978wp}, but the
focus was on constructing few-body models of reactions dominated by a
small number of few-body channels.  The interest at that time was
largely academic due to computational limitations. 
This work provides an alternate treatment of the method that appears
in \cite{Polyzou:1978wp}. It provides an alternate proof of the
optical theorem and uses time-dependent scattering theory to show that
the resulting approximate scattering wave functions in the retained
scattering channels agree with the exact scattering wave functions up
to, but not including, $N$-body correlations. The application to bound
systems and to relativistic systems, which is relevant for hadronic
reactions, is also discussed.

The representation of the Hamiltonian as a sum of channel Hamiltonians
is applicable to any nuclear Hamiltonian.  There are
no limitations on the choice of retained scattering channels, and the
method can be applied equally to theories with two-body and many-body
interactions. This is relevant because Hamiltonians generated using
effective field theory \cite{Epelbaum:2002vt}\cite{Epelbaum:1999dj} or
scattering equivalences
\cite{Ekstein:1960}\cite{Glockle:1990}\cite{Bogner:2006pc}\cite{polyzou_equiv}
generally have many-body interactions.

The next section summarizes the time-dependent formulation of
many-body scattering theory used in the subsequent sections.
Cluster expansions and their relation to expansions in
subsystem Hamiltonians are discussed in section three. The expansion
of the Hamiltonian as a sum of channel Hamiltonians 
is introduced in section four. The optical theorem for
any selected set of scattering channels is proved in section five.
Section six discusses a formulation of the time-independent treatment
of many-body scattering using the coupled equations of Bencze, Redish
and Sloan, which can be applied to the selected channel
Hamiltonian. The treatment of identical particles is discussed in
section seven. An example illustrating the structure of the dynamical
equations is given in section eight. Section nine discusses the
application to bound states. The treatment of relativistic reactions
is discussed in section ten. A summary and conclusion is given in
section eleven.

\section{Scattering Channels}
\label{sec:1}

\noindent The decomposition in \cite{Polyzou:1978wp} uses spectral
expansions of the Hamiltonians for all \textit{proper subsystems} as
input. These expansions involve complete sets of subsystem bound
states and scattering states. This section defines what is meant by a
scattering channel in the context of this work.

Let $H$ be the Hamiltonian for a system of $N$ particles with
short-range interactions. In general, the Hamiltonian $H$ will have
both two-body and many-body interactions. The notation
$a$ denotes a partition of the $N$ particles into $n_a$ non-empty
disjoint subsystems, labeled by $a_i$, and $H_{a_i}$ is the part of
$H$ involving only the particles in the $i^{th}$ subsystem of the
partition $a$.

In this work, scattering channels will always be associated with the
$N$-particle system. There is a scattering channel,
$\alpha$, associated with the partition $a$ if each subsystem
Hamiltonian, $H_{a_i}$, has a bound state or is a single particle Hamiltonian. 
A bound state associated with $H_{a_i}$ is denoted by
\[
\vert (E_i,j_i) \, \mathbf{p}_i, \mu_i\rangle \qquad \mbox{where} \qquad 1 \leq i \leq n_a .
\]
In this notation, $j_i$ is the total intrinsic angular momentum of the
$i^{th}$ bound state, $\mu_i$ is the magnetic quantum number of the
$i^{th}$ bound state, $\mathbf{p}_i$ is the total momentum of the
$i^{th}$ bound state, and
\[
E_i= \frac{\mathbf{p}_i^2}{2M_i} - e_{a_i}
\]
is the total kinetic energy minus the binding energy $e_{a_i}$ of
the $i^{th}$ bound subsystem ($M_i$ is the total mass of the $i^{th}$
bound subsystem). In general, for a given partition $a$ of the $N$
particles into $n_a$ subsystems, there may be one or more scattering
channels or zero channels associated with the partition $a$. The
notation ${\cal A}$ is used to denote the set of all scattering
channels of the $N$-body system, which by convention also includes the
one-body channels ($N$-body bound states). Except for the one-body
channels, the set ${\cal A}$ of scattering channels is determined by
the solution of \textit{proper subsystem problems}.

The notation discussed so far can be illustrated by considering the
subsystem Hamiltonians for a seven-particle system associated with the
partition $a = (135)(27)(46)$. There is a scattering channel
associated with this partition if each of the three subsystem Hamiltonians
can form bound states:
\[
a= \underbrace{(135)}_{{a_1}}\underbrace{(27)}_{{a_2}}\underbrace{(46)}_{{a_3}}
\qquad n_a=3 \qquad N=7=n_{a_1} + n_{a_2}+n_{a_3}
\]
\[
H_{a_1} = K_1 + K_3 + K_5 + V_{13} + V_{15} + V_{35} + V_{135}
\]
\[
H_{a_2} = K_2 + K_7 + V_{27}
\]
\[
H_{a_3} = K_4 + K_6 + V_{46}
\]
\[
H_{a_1} \vert (E_1,j_1) \, \mathbf{p}_1, \mu_1 \rangle = \left(\frac{\mathbf{p}_{1}^2}{2(m_1 + m_3 + m_5)} - e_{135}\right) \vert (E_1,j_1) \, \mathbf{p}_1, \mu_1 \rangle
\]
where $\mathbf{p}_1 = \mathbf{k}_1 + \mathbf{k}_3 + \mathbf{k}_5$, 
\[
H_{a_2} \vert (E_2,j_2) \, \mathbf{p}_2, \mu_2 \rangle = \left(\frac{\mathbf{p}_2^2}{2(m_2 + m_7)} - e_{27}\right) \vert (E_2,j_2) \, \mathbf{p}_2, \mu_2 \rangle
\]
where $\mathbf{p}_2 = \mathbf{k}_2 +\mathbf{k}_7$,
\[
H_{a_3} \vert (E_3,j_3) \, \mathbf{p}_3, \mu_3 \rangle = \left(\frac{\mathbf{p}_3^2}{2(m_4 + m_6)} - e_{46}\right) \vert (E_3,j_3) \, \mathbf{p}_3, \mu_3 \rangle
\]
where $\mathbf{p}_3 = \mathbf{k}_4 + \mathbf{k}_6$,
$\mathbf{k}_i$ are the single-particle momenta, $K_i$ are the
single-particle kinetic energies, $V_{ij}$ are two-body interactions,
$V_{135}$ is a three-body interaction, and
$e_{135}$, $e_{27}$ and $e_{46}$ are the binding
energies of the bound states.

For a given scattering channel $\alpha$, there are scattering states
associated with two different asymptotic conditions. The different
asymptotic conditions replace the initial conditions of the scattering
states with conditions that relate the
scattering states in the asymptotic past $(-)$ or asymptotic
future $(+)$ to states of non-interacting bound subsystems.
The scattering states, $\vert \Psi^{(\pm)}_{\alpha}
\rangle$, associated with the channel $\alpha$ are defined by strong
limits:
\beq
\lim_{t \to \pm \infty} \Vert \vert \Psi^{(\pm)}_{\alpha} \rangle - \sum_{\mu_1, \cdots, \mu_{n_a}} \int
e^{iHt} e^{-i H_a t} \otimes_{i=1}^{n_a} \vert (E_i, j_i) \, \mathbf{p}_i, \mu_i \rangle \, \phi_i(\mathbf{p}_i,\mu_i) \, d\mathbf{p}_i \Vert = 0 ,
\label{sc:1}
\eeq 
where the partition Hamiltonian, $H_a$, is the sum of subsystem Hamiltonians
\beq
H_a = \sum_{i=1}^{n_a} H_{a_i} \qquad \mbox{with} \qquad H_{a_i} \vert (E_i, j_i) \, \mathbf{p}_i,\mu_i \rangle = E_i \vert (E_i, j_i) \, \mathbf{p}_i, \mu_i \rangle
\label{sc:2}
\eeq
and satisfies
\beq
H_{a} \otimes_{i=1}^{n_a} \vert (E_i, j_i) \, \mathbf{p}_i, \mu_i \rangle = \left(\sum_{q=1}^{n_a} E_q\right) \otimes_{i=1}^{n_a} \vert (E_i, j_i) \, \mathbf{p}_i, \mu_i \rangle .
\label{sc:3}
\eeq
The operator $H_a$ is the part of the Hamiltonian with all of the
interactions between particles in the different clusters of the
partition $a$ turned off, and $\phi_i (\mathbf{p}_i,\mu_i)$ are wave
packets in the total momentum and magnetic quantum numbers of each
bound subsystem in the channel $\alpha$. The variables in the wave
packets are the experimentally detectable degrees of freedom (momentum
and spin polarization) of the bound subsystems.

The limit in (\ref{sc:1}) is a strong limit, and this means that the
integral over the wave packets must be computed {\it before} taking
the limit. If this is done in the correct order, then putting an extra
factor of $e^{\mp \epsilon t}$ and taking the limit as $\epsilon \to
0$ {\it after} performing the integral does not change the result. This
makes it possible to define
the limit using ``plane wave'' states where the $\epsilon \to 0$ limit
can be taken at the end of the calculation {\it after} integrating against
the wave packets. After including the factor of $e^{\mp \epsilon t}$,
the channel $\alpha$ scattering states
\beq
\vert \Psi^{(\pm)}_{\alpha} \rangle = \lim_{\epsilon \to 0} \sum_{\mu_1, \cdots, \mu_{n_a}} \int \vert \Psi^{(\pm)}_{\alpha}(\mathbf{p}_{1}, \mu_{1}, \cdots, \mathbf{p}_{n_a}, \mu_{n_a}) \rangle\prod_{i=1}^{n_a} \phi_{i} (\mathbf{p}_{i},\mu_{i}) \, d\mathbf{p}_i 
\label{sc:4}
\eeq
can be expressed in terms of the channel $\alpha$ ``plane wave'' scattering states defined by
\[
\vert \Psi^{(\pm)}_{\alpha}(\mathbf{p}_{1}, \mu_{1}, \cdots, \mathbf{p}_{n_a}, \mu_{n_a}) \rangle := \lim_{t \to \pm\infty} e^{iHt\mp \epsilon t}e^{-iH_at } \otimes_{i=1}^{n_a} \vert (E_i, j_i) \, \mathbf{p}_i, \mu_i \rangle =
\]
\[
\otimes_{i=1}^{n_a} \vert (E_i, j_i) \, \mathbf{p}_i, \mu_i\rangle + \lim_{t \to \pm\infty} \int_0^t  \frac{d}{dt} \left(e^{iHt\mp \epsilon t}e^{-iH_at }\right) \otimes_{i=1}^{n_a} \vert (E_i, j_i) \, \mathbf{p}_i, \mu_i \rangle \, dt =
\]
\[
\otimes_{i=1}^{n_a} \vert (E_i, j_i) \, \mathbf{p}_i, \mu_i\rangle
\]
\[
+ i \lim_{t \to \pm\infty} \int_0^t  e^{iHt\mp \epsilon t} \left(H \pm i \epsilon - H_a\right) e^{-iH_at } \otimes_{i=1}^{n_a} \vert (E_i, j_i) \, \mathbf{p}_i, \mu_i \rangle \, dt =
\]
\beq
\otimes_{i=1}^{n_a} \vert (E_i, j_i) \, \mathbf{p}_i, \mu_i \rangle + \left(E_{\alpha} - H \mp i\epsilon\right)^{-1} H^a \otimes_{i=1}^{n_a} \vert (E_i, j_i) \, \mathbf{p}_i, \mu_i \rangle
\label{sc:5} .
\eeq
The operator $H^a:= H-H_a$ is the sum of interactions between particles
in different clusters of $a$, and
\beq
E_{\alpha} = \sum_{q=1}^{n_a}\left(\frac{\mathbf{p}_q^2}{2M_q} - e_{a_q}\right)
\label{sc:6}
\eeq
is the total energy of the system ($M_q$ is the total mass of the
$q^{th}$ subsystem). The limit $\epsilon \to 0$ in (\ref{sc:5})
can only be taken after integrating against
products of wave packets which are functions of the momenta
and magnetic quantum numbers of each bound cluster.

The tensor product of the wave packets span a channel Hilbert space
${\cal H}_\alpha$. 
The operator, $\Phi_{\alpha}$, that maps the channel $\alpha$ Hilbert space
${\cal H}_\alpha$
to the $N$-body Hilbert space ${\cal H}$ is defined by
\beq
\Phi_{\alpha} \vert \phi_{o \alpha} \rangle := \sum_{\mu_1, \cdots, \mu_{n_a}} \int \otimes_{i=1}^{n_a} \vert (E_i, j_i) \, \mathbf{p}_i, \mu_i \rangle \, \phi_i(\mathbf{p}_i,\mu_i) \, d\mathbf{p}_i ,
\eeq
\label{sc:7}
where $\vert \phi_{o \alpha} \rangle \in {\cal H}_\alpha$ represents the product of wave packets given by 
\beq
\langle \mathbf{p}_1,\mu_1, \cdots , \mathbf{p}_{n_a}, \mu_{n_a} \vert \phi_{o \alpha} \rangle := \prod_{q=1}^{n_a} \phi_q (\mathbf{p}_q,\mu_q). 
\label{sc:8}
\eeq
The wave packets describe the experimentally accessible momentum and spin
distributions for the reaction. 
The mapping, $\Phi_{\alpha}:{\cal H}_{\alpha} \to {\cal H}$,
is called the channel injection operator, and it includes
the internal variables of the bound state wave functions for
each bound subsystem. In this two-Hilbert space notation, the channel $\alpha$ 
``plane wave'' scattering
states are expressed in terms of channel wave operators \cite{moller}:
\beq
\vert \Psi^{(\pm)}_{\alpha}(\mathbf{p}_1,\mu_1, \cdots , \mathbf{p}_{n_a},\mu_{n_a} ) \rangle =
\lim_{t \to \pm \infty} e^{iHt} e^{-iH_at}\Phi_{\alpha} \vert \phi_{o \alpha} \rangle =
\Omega^{(\pm)}(a) \Phi_{\alpha} \vert \phi_{o \alpha} \rangle ,
\label{sc:9}
\eeq
where
\beq
\Omega^{(\pm)}(a) :=\lim_{t \to \pm \infty} e^{iHt} e^{-iH_at} 
\label{sc:10}
\eeq
only makes sense as a strong limit applied to the normalizable vector
$\Phi_{\alpha} \vert \phi_{o \alpha} \rangle$. The advantage of the
notation in (\ref{sc:9}) is that it separates the part of the
scattering state that depends on the partition $a$ from the part that
depends on the associated scattering channel $\alpha$, and the operators 
$\Omega^{(\pm)}(a)$ act on the $N$-body Hilbert space.

The probability
amplitude density for a transition from an initial channel state
$\alpha$ to a final channel state $\beta$ (the scattering matrix) is
\[
\langle \Psi^{(+)}_{\beta}(\mathbf{p}'_1,\mu'_1,\cdots , \mathbf{p}'_{n_b},\mu'_{n_b}) \vert 
\Psi^{(-)}_{\alpha}(\mathbf{p}_1, \mu_1, \cdots, \mathbf{p}_{n_a},\mu_{n_a}) \rangle =
\]
\beq
\langle \beta, \mathbf{p}'_1, \mu'_1, \cdots, \mathbf{p}'_{n_b}, \mu'_{n_b} \vert S_{\beta\alpha} \vert \alpha, \mathbf{p}_1, \mu_1, \cdots, \mathbf{p}_{n_a}, \mu_{n_a} \rangle ,
\label{sc:11}
\eeq
where the channel scattering operator, $S_{\beta \alpha} := \Phi_\beta^{\dagger}\Omega^{(+) \dagger}(b) \Omega^{(-)}(a) \Phi_\alpha$, 
is used to express the scattering matrix in terms of the
non-interacting bound subsystems in the channels $\alpha$ and $\beta$
(the channel $\beta$ is associated with the partition $b$).
The channel scattering operator, $S_{\beta \alpha}$, is a mapping from ${\cal H}_\alpha$ to ${\cal H}_\beta$.  The channel Hilbert spaces are spaces of
square integrable functions of the experimentally observable degrees
of freedom in each scattering channel.

In a scattering process, the incoming $(-)$ states look like free
bound clusters long before the collision, and the outgoing $(+)$
states look like free bound clusters long after the collision. Since
there can be scattering from the channel $\alpha$ to the channel
$\beta$, the incoming and outgoing scattering states for different
channels with different asymptotic conditions ($\pm$) are not
orthogonal; however, the scattering states for different channels with
the same asymptotic condition ($\pm$) are orthogonal and complete if
the bound state channels are included. This assumes that the theory is
asymptotically complete, which is an assumption that the original
Hamiltonian is not pathological.

While the scattering matrix is the inner product of states satisfying
incoming $(-)$ and outgoing $(+)$ asymptotic conditions, it can be expressed in
terms of only the incoming scattering states
\[
\vert \Psi^{(-)}_{\alpha} (\mathbf{p}_1, \mu_1, \cdots, \mathbf{p}_{n_a}, \mu_{n_a}) \rangle.
\]
This is because $\Omega^{(+) \dagger}(a)$ and $\Omega^{(-)}(a)$ both involve limits of $e^{-iHt}$ with $t \to +\infty$. The resulting expression is
\[
\langle \beta, \mathbf{p}'_1, \mu'_1, \cdots, \mathbf{p}'_{n_b}, \mu'_{n_b} \vert S_{\beta \alpha} \vert \alpha, \mathbf{p}_1, \mu_1, \cdots, \mathbf{p}_{n_a}, \mu_{n_a} \rangle =
\]
\[
\delta_{\beta \alpha} \prod_i \delta(\mathbf{p}_i'- \mathbf{p}_i) \delta_{\mu_i'\mu_i} 
\]
\beq
- 2 \pi i \, \delta(E_{\beta}' - E_{\alpha}) 
\langle \mathbf{p}'_1, \mu'_1, \cdots, \mathbf{p}'_{n_b}, \mu'_{n_b} \vert \Phi_{\beta}^{\dagger} H^b \vert \Psi^{(-)}_{\alpha}(\mathbf{p}_1, \mu_1, \cdots, \mathbf{p}_{n_a},\mu_{n_a}) \rangle ,
\label{sc:12}
\eeq
where $H^b := H-H_b$ is the part of $H$ that only has interactions
between particles in different clusters of $b$. For short-range interactions,
the operator $H^b$
will vanish as the clusters of $b$ are asymptotically
separated. Equation (\ref{sc:12}) can be expressed in operator form
using the notation in (\ref{sc:9}):
\beq
S_{\beta\alpha} = I\delta_{\beta \alpha} - 2 \pi i \, \delta(E_{\beta}' - E_{\alpha}) \Phi_{\beta}^{\dagger} H^b \Omega^{(-)}(a) \Phi_{\alpha},
\label{sc:13}
\eeq
where
\beq
T^{\beta \alpha} := \Phi_{\beta}^{\dagger} H^b \Omega^{(-)}(a) \Phi_{\alpha}
\label{sc:14}
\eeq
is the right half-shell transition matrix element. The presence of the
energy conserving delta function $\delta(E_{\beta}' - E_{\alpha})$
ensures that the scattering matrix is only defined for
\textit{on-shell} values of the energy.

Assuming that the Hamiltonian commutes with the total linear momentum, the differential cross section for scattering from a 2-cluster channel
$\alpha$ to a general channel $\beta$ can be expressed in terms of the
above quantities as
\beq
\begin{aligned}
d\sigma = \frac{(2 \pi)^4}{\vert s \, \mathbf{v}_r \vert} \vert & \langle \mathbf{p}'_1, \mu'_1, \cdots, \mathbf{p}'_{n_b}, \mu'_{n_b} \Vert \Phi^{\dagger}_{\beta} H^b \Omega^{(-)}(a) \Phi_{\alpha} \Vert \mathbf{p}_1, \mu_1, \mathbf{p}_2, \mu_2 \rangle \vert^2 \\ & \times \, \delta(\sum_{j=1}^{n_b} E_j' - E_1 - E_2) \, \delta(\sum_{j=1}^{n_b}\mathbf{p}_j' - \mathbf{p}_1 - \mathbf{p}_2) \prod_{i=1}^{n_b} d\mathbf{p}_i' .
\end{aligned}
\label{sc:15}
\eeq
In the above expression, $\mathbf{v}_r$ is the relative speed of the
incoming pair of particles and $s = \prod_q \frac{1}{k_q!}$ is a
statistical normalization factor for identical bound states in the
final state, with $k_q$ denoting the number of identical bound states
of type $q$ in the final state. In (\ref{sc:15}), the $\Vert \cdots
\Vert$ indicates that a momentum conserving delta function has been
factored out of the expression so that
\[
\langle \mathbf{p}'_1, \mu'_1, \cdots, \mathbf{p}'_{n_b}, \mu'_{n_b} \vert \Phi^{\dagger}_{\beta} H^b \Omega^{(-)}(a) \Phi_{\alpha} \vert \mathbf{p}_1, \mu_1, \mathbf{p}_2, \mu_2 \rangle =
\]
\beq
\delta(\sum_{j=1}^{n_b}\mathbf{p}_j' - \mathbf{p}_1 - \mathbf{p}_2) \, \langle \mathbf{p}'_1, \mu'_1, \cdots, \mathbf{p}'_{n_b}, \mu'_{n_b} \Vert \Phi^{\dagger}_{\beta} H^b \Omega^{(-)}(a) \Phi_{\alpha} \Vert 
\mathbf{p}_1, \mu_1, \mathbf{p}_2, \mu_2 \rangle . 
\label{sc:16}
\eeq
The differential cross section in (\ref{sc:15}) contains several
independent variables, but in an experiment one chooses the
variables that will be measured and integrates over the remaining
variables in order to eliminate the delta functions. The differential
cross section is only defined for \textit{on-shell} matrix elements.

\section{Cluster Expansions}

\label{sec:2}

\noindent Cluster expansions play an important role in understanding
many-body reaction mechanisms and constructing approximations
\cite{Bazin}. For scattering, it is useful to keep track of operators
that satisfy or break translational invariance of subsystems. This
is because momentum-conserving delta functions are broken up by
short-range interactions. In this section, cluster expansions are
treated abstractly, and the abstraction provides a powerful framework
for managing cluster properties. Much of this section is based on
\cite{rota:1995}, see also \cite{Polyzou:1979wf}
\cite{Kowalski:1980cg}.

A partition $a$ of $N$ particles is an assignment of the $N$ particles
into distinct non-empty equivalence classes called clusters. The
following notation will be used is this paper:
\begin{itemize}
  
\item[] ${\cal P}_{N}$ is the set of all partitions for a system of $N$ particles;  
\item[] $n_a$ is the number of equivalence classes of $a$;
\item[] $a_i$ is the set of particles in the $i^{th}$ equivalence class of $a$;
\item[] $n_{a_i}$ is the number of particles in the $i^{th}$ equivalence class of $a$;
\item[] $i \sim_a j $ means that particles $i$ and $j$ are in the same equivalence class (cluster) of $a$;
\item[] $0:=\{(1)(2)\cdots (N)  \}$ is the unique $N$-cluster partition (each particle in a different class); 
\item[] $1:=\{(1 \cdots N)  \}$ is the unique $1$-cluster partition (all particles in the same class).
\end{itemize}
It follows from the definitions that
\beq
\sum_{i=1}^{n_a}n_{a_i} = N ,
\label{ce:1}
\eeq
and the number of classes $n_a$ for a given partition $a$ satisfies $1\leq n_a \leq N$. In order to demonstrate the use of partitions, here is a list of all the partitions of four particles: 
\begin{itemize}
  \item[] $(1234)$ is the unique 1-cluster partition;
  \item[] $(1)(234)$, $(2)(134)$, $(3)(124)$, $(4)(123)$, $(12)(34)$, $(13)(24)$, $(14)(23)$ are all of the 2-cluster partitions;
  \item[] $(12)(3)(4)$, $(13)(2)(4)$, $(14)(2)(3)$, $(23)(1)(4)$, $(24)(1)(3)$, $(34)(1)(2)$ are all of the 3-cluster partitions;
  \item[] $(1)(2)(3)(4)$ is the unique 4-cluster partition.
\end{itemize}

\noindent The above list exhausts all possible partitions of four
particles. The order of a particle within an equivalence class (or
cluster) does not matter. In this example, there are two types of
2-cluster partitions, $(ij)(kl)$ and $(i)(jkl)$, with each type involving 
partitions that are related by
permutations, and all of the 3-cluster partitions, $(ij)(k)(l)$, are
related by permutations.  Distinct partitions that are related by
permutations are called permutation equivalent partitions, and this is
important when dealing with systems of identical particles.

In what follows, partitions will be used to label parts of operators
that have no interactions between particles in different clusters
(equivalence classes) of the partition. The interactions between
particles in the same cluster of a partition are ``turned on'', while
the interactions between particles in different clusters are ``turned
off''. All of the partitions in the prior example satisfy
(\ref{ce:1}), and the notation so far can be illustrated by
considering the four-particle partition $a = (12)(3)(4)$:
\[
a= \underbrace{(12)}_{{a_1}}\underbrace{(3)}_{{a_2}}\underbrace{(4)}_{{a_3}}
\qquad n_a=3 \qquad N=4=n_{a_1} + n_{a_2}+n_{a_3} 
\] 
where this is the same notation that was used in the seven-particle
example from the previous section.

There is a natural partial ordering on the partitions $a$ and $b$ given by
\beq
a \supseteq b ,  
\label{ce:2}
\eeq
if every particle that is in the same $b$-equivalence class is in the
same $a$-equivalence class ($i \sim_b j \to i \sim_a j$).

The partial ordering for a system of particles is illustrated by the following
example:
\[
\left .
\begin{array}{cc}
 a = (12)(34) \\
 b = (1)(2)(34) \\
 c = (123)(4) \\
\end{array}
\right \} 
=> a \supseteq b, c \nsupseteq b .
\]
In this example, there is a partial ordering on partitions $a$ and $b$
because particles 3 and 4 are in the same cluster in both partitions (with 
partition $a$ only including the additional interaction between particles
1 and 2). There is no ordering on partitions $c$ and $b$ because 
particles 3 and 4 are not in the same cluster in both partitions.

For two partitions, $a$ and $b$, the union $a \cup b$ is the least
upper bound of $a$ and $b$ with respect to the partial ordering, and
the intersection $a \cap b$ is the greatest lower bound of $a$ and $b$
with respect to the partial ordering. The union and intersection are
formally defined by

\begin{itemize}
\item[] $a\cup b$: $(a\cup b) \supseteq a$,  $(a\cup b) \supseteq b$, and if $c \supseteq a$, $c \supseteq b$ then $c \supseteq (a \cup b)$

\item[] $a\cap b$: $a \supseteq (a\cap b)$,  $b \supseteq (a\cap b)$, and if $a \supseteq c$, $b \supseteq c$, then $(a \cap b) \supseteq c$ .
\end{itemize}
The union and intersection for a system of
particles is illustrated by the following example:
\[
\left .
\begin{array}{cc}
 a = (123)(4567)(89) \\
 b = (1234)(567)(89) \\
\end{array}
\right \} 
=> a \cup b = (1234567)(89), a \cap b = (123)(4)(567)(89) .
\]
This example demonstrates how the union and intersection of the
partitions $a$ and $b$ can be used to construct new partitions that
set a least upper bound and greatest lower bound, respectively, on the
partial ordering of the partitions $a$ and $b$. It can be seen from
the definitions and the above example that every partition $a$
satisfies $1 \supseteq a \supseteq 0$.

This structure for the partial orderings on partitions is called a
partition lattice. Some important tools, that will be utilized in what
follows, are the incidence function and its inverse. These functions
are also called the Zeta and M\"obius functions on the partition lattice, respectively. The Zeta
function is defined in \cite{rota:1995}\cite{Polyzou:1979wf}\cite{Kowalski:1980cg}
as
\beq
\Delta_{a \supseteq b} :=
\left \{
\begin{array}{cc}
 1 & \mbox{if} \; \; a \supseteq  b \\
 0 & \mbox{if} \; \; a \not\supseteq  b\\
\end{array}
\right .  .
\eeq
Since this is upper triangular with 1's on the diagonal, it
necessarily has an inverse given by
\beq
\Delta^{-1}_{a \supseteq b} :=
\left \{
\begin{array}{cc}
(-)^{n_a}\prod_{i=1}^{n_a} (-)^{n_{b_i}}(n_{b_i}-1)! & \mbox{if} \; \; a \supseteq  b \\
 0 & \mbox{if} \; \; a \not\supseteq  b\\
\end{array} ,
\right . 
\eeq
where $n_{b_i}$ is the number of clusters of $b$ in the $i^{th}$
cluster of $a$. Note that both the Zeta function, $\Delta_{a
\supseteq b}$, and M\"obius function, $\Delta^{-1}_{a \supseteq b}$,
vanish when $a \not\supseteq b$. The Zeta and M\"obius functions are
matrix operators that satisfy
\beq
\sum_{c \in {\cal{P}}_{N}}{\Delta^{-1}_{a \supseteq c} \Delta_{c \supseteq b}} = \delta_{ab} \qquad \mbox{and} \qquad \sum_{a} \delta_{ab} = 1 \qquad \mbox{for} \qquad a \supseteq b ,
\eeq
which follows from the definitions of the Zeta and M\"obius functions.

Partitions can be used to classify operators on the $N$-particle
Hilbert space. The starting assumption is that the Hamiltonian $H$ is
translationally invariant and commutes with the total momentum
operator. Each cluster $a_i$ of the $N$-body system represents a
subsystem, and the total momentum $\mathbf{p}_{a_i}$ of the particles
in the cluster $a_i$ is
\beq
\mathbf{p}_{a_i} := \sum_{j \in a_i} \mathbf{k}_j .
\label{ce:3}
\eeq
The quantity $\mathbf{k}_j$ denotes the single-particle momenta, and
the total momentum $\mathbf{p}_{a_i}$ is the generator of translations
of the subsystem of particles in the cluster $a_i$. The operator that
independently translates each cluster $a_i$ of the partition $a$ by a
vector $\mathbf{x}_i$ is
\beq
T_a(\mathbf{x}_1, \cdots , \mathbf{x}_{n_a}) := e^{i \sum_{i=1}^{n_a} \mathbf{x}_i \cdot \mathbf{p}_{a_i}} .
\label{ce:4}
\eeq
This is a $3n_a$ parameter unitary group of translations. An
operator $O$ that commutes with $T_a(\mathbf{x}_1, \cdots ,
\mathbf{x}_{n_a})$, 
and satisfies
\beq
[O,T_a(\mathbf{x}_1, \cdots , \mathbf{x}_{n_a})] = 0 
\label{ce:5}
\eeq
for all $\mathbf{x}_i$, is called an $a$-invariant operator.

For any partition $a$ of the particles into non-empty disjoint
subsystems, a general operator can be expressed as the sum of an
operator that commutes with $T_a(\mathbf{x}_1, \cdots,
\mathbf{x}_{n_a})$ and a remainder. This is represented by the
notation

\beq
O = O_a + O^a ,
\label{ce:6}
\eeq
where $O_a$ is the $a$-invariant part of $O$ and $O^a:= O-O_a$ is the
remainder that breaks the $T_a(\mathbf{x}_1, \cdots,
\mathbf{x}_{n_a})$ translational invariance. It follows from the definitions that 
\beq
\Vert
T_a^{\dagger} (\mathbf{x}_1, \cdots, \mathbf{x}_{n_a}) \left(O - O_a\right) T_a(\mathbf{x}_1, \cdots, \mathbf{x}_{n_a}) \vert \psi \rangle \Vert = \Vert O^a \, T_a(\mathbf{x}_1, \cdots, \mathbf{x}_{n_a}) \vert \psi \rangle \Vert 
\label{ce:7}
\eeq
for the state $\vert \psi \rangle$. For an operator $O$ that is
overall translationally invariant, $O^a$ involves operators that only
contain interactions between particles in different clusters of the
partition $a$. If these are all short-range interactions, then as all
of the clusters of $a$ are asymptotically separated $O^a$ should
vanish. A mathematical formulation of this condition is
\beq
\lim_{\vert \mathbf{x}_i -\mathbf{x}_j \vert \to \infty } \Vert O^a \, T_a(\mathbf{x}_1, \cdots, \mathbf{x}_{n_a}) \vert \psi \rangle \Vert = 0 .
\label{ce:8}
\eeq
A many-body operator $O$ that can be decomposed as $O=O_a+O^a$
with $O_a$ satisfying (\ref{ce:5}) and $O^a$ satisfying (\ref{ce:8})
will be called a
translationally fibered operator. The types of operators considered
in this work are interactions, projections on bound subsystems,
resolvents of the form in (\ref{sc:5}), wave operators, and time
evolution operators. For suitable short-range interactions, limits of
the form (\ref{ce:8}) are expected to vanish. For translationally
fibered operators, it follows from (\ref{ce:5}) and (\ref{ce:8}) that
$O_a$ can be constructed from $O$ using
\beq
O_a = \lim_{\vert \mathbf{x}_i - \mathbf{x}_j \vert \to \infty} T_a^{\dagger} (\mathbf{x}_1, \cdots, \mathbf{x}_{n_a}) \, O \, T_a(\mathbf{x}_1, \cdots, \mathbf{x}_{n_a}) ,
\label{ce:9}
\eeq
which shows that $O_a$ can be obtained by asymptotically separating
the different clusters of the partition $a$. This notation can
be illustrated by considering the four-particle Hamiltonian associated
with the partition $a = (1)(2)(34)$:
\[
H = \underbrace{K_1 + K_2 + K_3 + K_4 + V_{34}}_{H_{(1)(2)(34)}} +
\]
\[
\underbrace{V_{12} + V_{13} + V_{14} + V_{23} + V_{24} + V_{123} + V_{124} + V_{134} + V_{234} + V_{1234}}_{H^{(1)(2)(34)}},  
\]
where $K_i$ are the single-particle kinetic energies, $V_{ij}$ are
two-body interactions, $V_{ijk}$ are three-body interactions, and
$V_{1234}$ is a four-body interaction. In this example,
$H^{(1)(2)(34)}$ vanishes when all of the clusters of $a = (1)(2)(34)$ are
asymptotically separated, and one is left with the $a$-invariant
operator $H_{(1)(2)(34)}$.  The interactions in $H^{(1)(2)(34)}$ all involve
particles in {\it different} clusters of $a$.

For $b \supseteq a$, $T_b(\mathbf{x}_1, \cdots, \mathbf{x}_{n_b})$ is
a subgroup of $T_a(\mathbf{x}_1, \cdots, \mathbf{x}_{n_a})$, and it
follows that
\beq
O_a = \lim_{\vert \mathbf{x}_i -\mathbf{x}_j \vert \to \infty} T^{\dagger}_b(\mathbf{x}_1, \cdots, \mathbf{x}_{n_b}) \, O_a \, T_b(\mathbf{x}_1, \cdots, \mathbf{x}_{n_b}) \qquad \mbox{for} \qquad b \supseteq a .
\label{ce:10}
\eeq
This means that $O_a$ is invariant with respect to translations that
separate the clusters of $b$ when $b \supseteq a$. As an example,
consider the partitions $a = (1)(2)(34)$ and $b = (12)(34)$ which
satisfy $b \supseteq a$. The operator $H_{(1)(2)(34)}$ from the
previous example is invariant with respect to translations that
asymptotically separate the clusters of $b = (12)(34)$ because these
translations do not separate particles 3 and 4.

On the other hand, if $b \not\supseteq a$, then $O_a$ has the decomposition
\beq
O_a = (O_{a})_b + (O_a)^b = O_{a \cap b} + O_a^b .
\label{ce:11}
\eeq
In this decomposition, $O_{a \cap b}$ is invariant with respect to
both $T_a(\mathbf{x}_1, \cdots, \mathbf{x}_{n_a})$ and
$T_b(\mathbf{x}_1, \cdots, \mathbf{x}_{n_b})$, and $O_a^b$ vanishes as
the clusters of the partition $b$ are asymptotically separated. Using
(\ref{ce:8}),
\beq
\lim_{\vert \mathbf{x}_i -\mathbf{x}_j \vert \to \infty } \Vert O^b_a \, T_b(\mathbf{x}_1, \cdots, \mathbf{x}_{n_b}) \vert \psi \rangle \Vert = 0 ,
\label{ce:12}
\eeq
which means that 
\beq
O_{a \cap b} = \lim_{\vert \mathbf{x}_i - \mathbf{x}_j \vert \to \infty } T^{\dagger}_b(\mathbf{x}_1, \cdots, \mathbf{x}_{n_b}) \, O_a \, T_b(\mathbf{x}_1, \cdots, \mathbf{x}_{n_b}) \qquad \mbox{for} \qquad b \not\supseteq a .
\label{ce:13}
\eeq
Therefore, $O_{a \cap b}$ can be obtained from $O_a$ by asymptotically
separating the clusters of $b$ when $b \not\supseteq a$. As an
example, consider the partitions $a = (1)(2)(34)$ and $b = (123)(4)$
which satisfy $b \not\supseteq a$. The operator $H_{(1)(2)(34)}$ from
the prior examples can be written as
\[
H_{(1)(2)(34)} = \underbrace{K_1 + K_2 + K_3 + K_4}_{H_{(1)(2)(34) \cap (123)(4)}} + \underbrace{V_{34}}_{H_{(1)(2)(34)}^{(123)(4)}} .
\]
In this example, $H_{(1)(2)(34)}^{(123)(4)}$ vanishes when the
clusters of $b = (123)(4)$ are asymptotically separated, and one is
left with $H_{(1)(2)(34) \cap (123)(4)}= H_{(1)(2)(3)(4)}$. It should be 
noted that a translationally fibered
operator can always be decomposed as
(\ref{ce:11}), but the term $O_{a}^{b}$ vanishes whenever $b \supseteq
a$ (this is because $T_b(\mathbf{x}_1, \cdots, \mathbf{x}_{n_b})$ is a
subgroup of $T_a(\mathbf{x}_1, \cdots, \mathbf{x}_{n_a})$).

For applications, it is useful to define $[O]_a$, the {\it
$a$-connected part} of $O$, by the conditions
\beq
[[O]_a, T_a(\mathbf{x}_1, \cdots, \mathbf{x}_{n_a})] = 0 \qquad \mbox{and} \qquad ([O]_a)_b = 0  \qquad \mbox{for} \qquad b \not\supseteq a .
\label{ce:14}
\eeq
This means that the operator $[O]_a$ is invariant with respect to the translations
$T_a(\mathbf{x}_1, \cdots, \mathbf{x}_{n_a})$, but it is not invariant
with respect to the translations $T_b(\mathbf{x}_1,
\cdots, \mathbf{x}_{n_b})$ when $b \not\supseteq
a$ (these translations necessarily break up at least one of the
clusters of $a$). For $b \supseteq a$, it follows that the
$b$-invariant part of $O$ is a sum of the $a$-connected parts of $O$
that commute with $T_b(\mathbf{x}_1, \cdots, \mathbf{x}_{n_b})$. This
means that
\beq
O_b = \sum_{ b \supseteq a} [O]_a = \sum_{a \in {\cal P}_{N}} \Delta_{b \supseteq a} [O]_a ,
\label{ce:15}
\eeq
where $O_b$ is expressed in terms of the Zeta function on the partition lattice. This expression can be inverted using the M\"obius function on the partition lattice
\beq
[O]_b = \sum_{a \in {\cal P}_{N}} \Delta^{-1}_{b \supseteq a} O_a .
\label{ce:16}
\eeq
It follows from (\ref{ce:15}) and (\ref{ce:16}) that the Zeta and
M\"obius functions on the partition lattice provide a direct relation
between the operators $O_b$ and $[O]_b$. Additionally, an operator $O$ is said to be
completely connected if $O=[O]_1$, and this means that $O$ vanishes in
the limit that any pair of particles are asymptotically separated. It
follows from these expressions that
\beq
[O]_1 = \sum_{a \in {\cal P}_{N}} \Delta^{-1}_{1 \supseteq a} O_a = \Delta^{-1}_{1 \supseteq 1} O_1 + \sum_{a \in {\cal P}_{N}'} \Delta^{-1}_{1 \supseteq a} O_a = O + \sum_{a \in {\cal P}_{N}'} \Delta^{-1}_{1 \supseteq a} O_a ,
\label{ce:17}
\eeq
where ${\cal P}_{N}'$ is the set of all partitions of $N$ particles
excluding the 1-cluster partition. This means $O$ has the
decomposition
\beq
O = [O]_1 - \sum_{a \in {\cal P}_{N}'} \Delta^{-1}_{1 \supseteq a} O_a .
\label{ce:18} 
\eeq
It is useful to define the coefficient appearing in (\ref{ce:18}) as
\beq
{\cal C}_a := - \Delta^{-1}_{1\supseteq a} = (-)^{n_a} (n_a-1)! \qquad \mbox{with} \qquad \sum_{a \in {\cal P}_{N}'} {\cal C}_a = 1 ,
\label{ce:19}
\eeq
which is a combinatoric factor that ensures that the decomposition has the correct overall counting. The sum in (\ref{ce:19}) follows because
\[
\sum_{a \in {\cal P}_{N}'} \Delta^{-1}_{1 \supseteq a} = 
-\Delta^{-1}_{1 \supseteq 1} + \sum_a \Delta^{-1}_{1 \supseteq a} =
-\Delta^{-1}_{1 \supseteq 1} + \sum_a \Delta^{-1}_{1 \supseteq a} \underbrace{\Delta_{a \supseteq 0}}_{= 1}  = 
-1 + 0 .
\]
An important consequence of the invertibility of the incidence matrix
is that a general translationally fibered operator can be expressed in
two equivalent ways:
\beq
O = \sum_{a \in {\cal P}_{N}} [O]_a = [O]_1 + \sum_{a \in {\cal P}_{N}'} {\cal C}_a O_a .
\label{ce:20}
\eeq
The first sum is the cluster expansion of $O$, and the second sum is
the operator decomposition of $O$ in terms of proper subsystem operators.
This is a generalization of the
linked cluster theorem for identical particles.
The cluster expansion is a sum over
the $a$-connected parts of $O$, and the operator decomposition
consists of the completely connected part of $O$ (the $N$-body
interaction) and a linear combination of the $a$-invariant parts of
$O$. For an operator like a Hamiltonian, the cluster expansion is a
linear combination of interactions, while the sum over the
$a$-invariant parts of the Hamiltonian in the operator decomposition
corresponds to a linear combination of proper subsystem
Hamiltonians. As an example, a three-body Hamiltonian with
two-body and three-body interactions can be expressed as a cluster expansion or
as a sum of proper subsystem Hamiltonians:
\[
H = \underbrace{K_1 + K_2 + K_3}_{[H]_0 = [H]_{(1)(2)(3)}} +
\underbrace{V_{12}}_{[H]_{(12)(3)}} +
\underbrace{V_{13}}_{[H]_{(13)(2)}} + \underbrace{V_{23}}_{[H]_{(23)(1)}} + \underbrace{V_{123}}_{[H]_1 = [H]_{(123)}} =
\]
\[
-2 \underbrace{\left(K_1 + K_2 + K_3\right)}_{H_0 = H_{(1)(2)(3)}=H_1+H_2+H_3} + \underbrace{K_1 + K_2 + K_3 + V_{12}}_{H_{(12)(3)}=H_{12}+H_3}
\]
\[
+ \underbrace{K_1 + K_2 + K_3 + V_{13}}_{H_{(13)(2)}=H_{13}+H_2}
+ \underbrace{K_1 + K_2 + K_3 + V_{23}}_{H_{(23)(1)}=H_{23}+H_1} + \underbrace{V_{123}}_{[H]_1 = [H]_{(123)}} 
\]
where $K_i$ are the single-particle kinetic energies, $V_{ij}$ are
two-body interactions, $V_{123}$ is the three-body interaction
(completely connected part), and $H_{(ij)(k)}$ is a sum of subsystem
Hamiltonians. The coefficients in the second and third lines are the
combinatoric factors ${\cal C}_a$, and they ensure that the cluster
expansion and operator decomposition are equal to one another.
In this example, the three kinetic energy terms appear in each of the
three 2-cluster partition Hamiltonians, and the $(-2)$ in front of the 3-cluster
Hamiltonian corrects for this overcounting of the three kinetic energy terms. This
example demonstrates how the operator decomposition in (\ref{ce:20}) is used to express the cluster decomposition of the $N$-body Hamiltonian in terms of
a linear combination of subsystem Hamiltonians. The decomposition in
(\ref{ce:20}) is useful for identifying the parts of the Hamiltonian that are responsible for
the different channel asymptotic states in a manner that treats all
scattering channels
democratically.

In general, if $A$ and $B$ are bounded operators, then
\[
\Vert T_a^{\dagger} AB T_a \vert \psi \rangle \Vert = \Vert T_a^{\dagger} (A_a + A^a)(B_a + B^a) T_a \vert \psi \rangle \Vert \leq
\]
\[
\Vert A_a B_a \vert \psi \rangle \Vert + \Vert T_a^{\dagger} A^a T_a B_a \vert \psi \rangle \Vert +
\Vert A_a T_a^{\dagger} B^a T_a \vert \psi \rangle \Vert + \Vert T_a^{\dagger} A^a T_a T_a^{\dagger} B^a T_a \vert \psi \rangle \Vert \leq
\]
\[
\Vert A_a B_a \vert \psi \rangle \Vert + \Vert T_a^{\dagger} A^a T_a B_a \vert \psi \rangle \Vert +
\Vert A_a \Vert \Vert T_a^{\dagger} B^a T_a \vert \psi \rangle \Vert + \Vert A^a T_a \Vert \Vert T_a^{\dagger} B^a T_a \vert \psi \rangle \Vert \leq
\]
\[
\Vert A_a B_a \vert \psi \rangle \Vert + \Vert A^a T_a B_a \vert \psi \rangle \Vert + \Vert A_a \Vert \Vert B^a T_a \vert \psi \rangle \Vert + \Vert A^a T_a \Vert \Vert B^a T_a \vert \psi \rangle \Vert .
\]
If $A$ and $B$ are translationally fibered operators, then the last
three terms in the fourth line vanish in the limit that all of the
clusters of $a$ are asymptotically separated. This means that
translationally fibered operators satisfy
\beq
(AB)_a = A_a B_a .
\label{ce:21}
\eeq
It follows from (\ref{ce:18}) and (\ref{ce:19}) that 
\[
\sum_{a \in {\cal P}_{N}'} {\cal C}_a A_a B_a = AB - [AB]_1 = \left(\sum_{a \in {\cal P}_{N}'} {\cal C}_a A_a + [A]_1\right) \left(B_a + B^a\right) - [AB]_1 =
\]
\[
\sum_{a \in {\cal P}_{N}'} {\cal C}_a A_a B_a + \sum_{a \in {\cal P}_{N}'} {\cal C}_a A_a B^a +[A]_1 B - [AB]_1 .
\]
Canceling $\sum_{a \in {\cal P}_{N}'} {\cal C}_a A_a B_a$ on both sides
of this equation gives
\beq
\sum_{a \in {\cal P}_{N}'} {\cal C}_a A_a B^a  = -[A]_1 B + [A B]_1 ,
\label{ce:22}
\eeq
where the terms on the right are connected.
This means that sums of the form
\beq
\sum_{a \in {\cal P}_{N}'} {\cal C}_a A_a B^a
\label{ce:23}
\eeq
are either 0 or connected.

For an $N$-particle system, the dynamics is given by the unitary time
evolution operator $U(t)=e^{-iHt}$ where $H$ is the $N$-particle
Hamiltonian.  By turning off the interactions between particles in different clusters of the partition $a$, $H$ becomes $H_a$, and this is the infinitesimal
generator of time translation, $U_a(t)$, of non-interacting clusters of $a$.
In general, due to the kinetic energy terms,
Hamiltonians are not bounded operators, but Hunziker in
\cite{hunziker} proved that $U(t)$ satisfies (\ref{ce:8}) for
Hamiltonians with square integrable interactions (i.e., the
Hamiltonian is a translationally fibered operator). This means that
\beq
\lim_{\vert \mathbf{x}_i -\mathbf{x}_j \vert \to \infty } \Vert U^a(t) \, T_a(\mathbf{x}_1, \cdots,\mathbf{x}_{n_a}) \vert \psi \rangle \Vert = 0 .
\label{ce:24}
\eeq
He also proved that, under the same assumptions, the wave operators are translationally fibered. In everything that follows, it will be assumed that these properties
are satisfied.

\section{Channel Decomposition}
\label{sec:3}

\noindent The set of scattering channels ${\cal A}$ can be separated
into disjoint sets of channels; a selected set ${\cal A}_1$ and a
remainder ${\cal A}_2$ \cite{Polyzou:1978wp}. This section provides
the construction of a representation of a general $N$-particle Hamiltonian
$H$ as a sum of a Hamiltonian $H_{{\cal A}_1}$ that depends on the channels
${\cal A}_1$ and a Hamiltonian $H_{{\cal A}_2}$ that depends on the
complementary set of channels ${\cal A}_2$.

For a general many-body Hamiltonian, the exact spectral decomposition
has the form
\beq
H = \sum_{\alpha \in {\cal A}} \vert \psi^{(-)}_\alpha \rangle \langle \psi^{(-)}_\alpha \vert H =
\sum_{\alpha \in {\cal A}} P^{(-)}_{\alpha} H \qquad \mbox{with } \qquad I = \sum_{\alpha \in {\cal A}} P^{(-)}_{\alpha} ,
\label{hd:1}
\eeq
where 
\beq
P^{(-)}_{\alpha} := \Omega^{(-)}(a) \Phi_{\alpha} \Phi_{\alpha}^{\dagger} \Omega^{(-) \dagger}(a)  
\label{hd:2}
\eeq
is the orthogonal projection on the subspace spanned by the $(-)$
scattering states in the channel $\alpha$. By convention, the channel
sum in (\ref{hd:1}) includes the one-body channels ($N$-body bound
states). The notation in (\ref{hd:2}) is shorthand for
\[
P^{(-)}_{\alpha} H :=
\sum_{\mu_1, \cdots, \mu_{n_a}} \int d\mathbf{p}_1 \cdots d\mathbf{p}_{n_a} \, \vert (\alpha, \mathbf{p}_1, \mu_1, \cdots, \mathbf{p}_{n_a}, \mu_{n_a})^{(-)} \rangle E_\alpha \times
\]
\[
\langle (\alpha, \mathbf{p}_1, \mu_1, \cdots, \mathbf{p}_{n_a}, \mu_{n_a})^{(-)} \vert ,
\]
where $E_{\alpha}$ is the total energy eigenvalue defined in (\ref{sc:6}). Both
the Hamiltonian $H$ and the channel projection operator
$P^{(-)}_{\alpha}$ have cluster expansions of the form in (\ref{ce:20}):
\beq
H = \sum_{b \in {\cal P}_N}[H]_b = [H]_1 + \sum_{b \in {\cal P}_{N}'} {\cal C}_b H_b
\qquad \mbox{and} \qquad P^{(-)}_{\alpha} = \sum_{b \supseteq a} [P^{(-)}_{\alpha}]_b ,
\label{hd:3}
\eeq
where $H_b$ is the infinitesimal generator of $U_b(t)$. The product of
the Hamiltonian $H$ and the channel projection operator
$P^{(-)}_{\alpha}$ also has a cluster expansion. The
expansions of these operators, using (\ref{ce:21}), gives
\beq
H = \sum_{\alpha \in {\cal A}} \sum_{\{b \in {\cal P}_N \vert b \supseteq a\}} [P^{(-)}_{\alpha} H]_b ,
\label{hd:4}
\eeq
where $[P^{(-)}_{\alpha} H]_b$ is the $b$-connected part of
$P^{(-)}_{\alpha} H = \Omega^{(-)}(a) \Phi_{\alpha}
\Phi_{\alpha}^{\dagger} \Omega^{(-) \dagger}(a) H$. Using
(\ref{ce:18}) and (\ref{ce:19}), (\ref{hd:4}) can be decomposed into a
linear combination of its $b$-invariant parts and a completely
connected part:
\beq
H = \sum_{\alpha \in {\cal A}} \left([P^{(-)}_{\alpha} H]_1 + \sum_{\{b \in {\cal P}_{N}' \vert b \supseteq a\}} {\cal C}_b (P^{(-)}_{\alpha})_b H_b \right) ,
\label{hd:5}
\eeq
where $[P^{(-)}_{\alpha} H]_1$ is the completely connected part of the
product $P^{(-)}_{\alpha} H$ and $a$ is the partition associated with
the bound clusters of the channel $\alpha$. For $b \not\supseteq a$,
translating the clusters of $b$ will separate particles in at least
one of the bound clusters of $a$ in the channel $\alpha$. Therefore,
\beq
(\Phi_\alpha \Phi^{\dagger}_{\alpha})_b = ([\Phi_\alpha \Phi^{\dagger}_{\alpha}]_a)_b = 0 \qquad \mbox{for} \qquad b \not \supseteq a ,
\label{hd:6}
\eeq
which means that
\beq
(P^{(-)}_{\alpha})_b = 0 \qquad \mbox{for} \qquad b \not\supseteq a .
\label{hd:7}
\eeq
This means that the sum over the $b$-invariant parts of $P^{(-)}_{\alpha}
H$ is zero when $b \not\supseteq a$.

Since the interactions between particles in different clusters of $a$
that are in the same clusters of $b$ are short-range, the wave
operators satisfy a chain rule
\cite{Kato}\cite{bencze:1976} that allows successive interactions to
be turned on. This same result follows from the analysis in
\cite{hunziker}. For any $b$ satisfying $b \supseteq a$, the chain
rule for wave operators gives
\[
\Omega^{(-)}(a) \Phi_{\alpha} = \lim_{t \to -\infty}e^{iHt} e^{-iH_at} \Phi_{\alpha} =
\]
\[
\lim_{t \to -\infty} e^{iHt} \underbrace{e^{-iH_bt} e^{iH_bt}}_{I} e^{-iH_at} \Phi_{\alpha} = \lim_{t \to -\infty} e^{iHt}e^{-iH_bt} (\Omega^{(-)}(a))_b \Phi_{\alpha} =
\]
\beq
\Omega^{(-)}(b) (\Omega^{(-)}(a))_b \Phi_{\alpha} .
\label{hd:8}
\eeq
Here $(\Omega^{(-)}(a))_b \Phi_{\alpha}$ replaces $\Phi_{\alpha}$ when computing $\Omega^{(-)}(b)$. Since $\Omega^{(-)}(b)$ turns on the interactions between particles in different clusters of the partition $b$, then when $b \supseteq a$ one can write
\beq
(\Omega^{(-)}(a))_b \Phi_{\alpha} = 
((\Omega^{(-)}(b))_b (\Omega^{(-)}(a))_b \Phi_{\alpha})_b .
\label{hd:9}
\eeq
This shows that $(\Omega^{(-)}(b))_b$ acts like the identity on 
$(\Omega^{(-)}(a))_b \Phi_{\alpha}$, and 
this means that
\[
(\Omega^{(-)}(a))_b \Phi_{\alpha} = ((\Omega^{(-)}(a))_b
\Phi_{\alpha})_b
\]
is the $b$-invariant part of $\Omega^{(-)}(a)
\Phi_{\alpha}$ for $b \supseteq a$.
The important point is that this
involves solutions of the scattering problem in the channel $\alpha$
for the Hamiltonian $H_b = \sum_{i=1}^{n_b}H_{b_i}$, which is a sum of
proper subsystem Hamiltonians $H_{b_i}$. It also implies that every
$H_b$ for $b\supseteq a$ has channel $\alpha$ scattering states. The
operator $(\Omega^{(-)}(a))_b \Phi_\alpha \Phi^{\dagger}_{\alpha}
(\Omega^{(-) \dagger}(a))_b$ is the part of the exact spectral
projection that remains after turning off the interactions between
particles in different clusters of $b$ (there are still remaining
interactions between the asymptotically bound subsystems in the different clusters of
$a$ that are in the same clusters of $b$).

It follows that the exact projection of the Hamiltonian on the channel
$\alpha$ subspace has the decomposition
\[
P^{(-)}_{\alpha} H = [P^{(-)}_{\alpha} H]_1 + \sum_{\{b \in {\cal P}_{N}' \vert b \supseteq a\}}{\cal C}_b (P^{(-)}_{\alpha})_b H_b =
\]
\beq
[P^{(-)}_{\alpha} H]_1 + \sum_{\{b \in {\cal P}_{N}' \vert b \supseteq a\}} {\cal C}_b(\Omega^{(-)}(a))_b \Phi_\alpha \Phi^{\dagger}_{\alpha} (\Omega^{(-) \dagger}(a))_b H_b .
\label{hd:10}
\eeq
Up to this point, everything is exact. The cluster properties imply
that the terms $(\Omega^{(-)}(a))_b \Phi_\alpha
\Phi^{\dagger}_{\alpha} (\Omega^{(-) \dagger}(a))_b H_b$ in
(\ref{hd:10}), for $b \in {\cal P}_{N}'$, can be computed using only
\textit{proper subsystem solutions}. The assumed asymptotic
completeness implies that the sum over all channel projectors is the
identity
\beq
I = \sum_{\alpha \in {\cal A}} P^{(-)}_{\alpha}.
\label{hd:11}
\eeq
It follows that the Hamiltonian can be expressed as
\[
H = [H]_1 + \sum_{b \in {\cal P}_{N}'} {\cal C}_b H_b =
\label{hd:12}
\]
\[
\sum_{\alpha\in {\cal A}} \left([P^{(-)}_{\alpha}H]_1 + \sum_{\{b \in {\cal P}_{N}' \vert b \supseteq a\}} {\cal C}_b (P^{(-)}_{\alpha})_b H_b\right) =
\]
\beq
\sum_{\alpha\in {\cal A}} \left([P^{(-)}_{\alpha} H]_1 + \sum_{\{b \in {\cal P}_{N}' \vert b \supseteq a\}} {\cal C}_b (\Omega^{(-)}(a))_b \Phi_\alpha \Phi^{\dagger}_{\alpha} (\Omega^{(-) \dagger}(a))_b H_b\right) .
\label{hd:13}
\eeq
Comparing these expressions, the completely connected parts on both sides
of (\ref{hd:13}) must be the same and are given by
\beq
[H]_1 = \sum_{\alpha \in {\cal A}}[P^{(-)}_{\alpha}H]_1 .
\label{hd:14}
\eeq
This means that they add up to zero if $H$ does not have an
$N$-body interaction.

The next step is to introduce the channel decomposition, and this is
the main result of \cite{Polyzou:1978wp}.  This is done by decomposing the
collection of channels into two disjoint sets, ${\cal A} = {\cal
  A}_1\cup {\cal A}_2$, where ${\cal A}_1$ is a selected
set of scattering channels
and ${\cal A}_2$ represents the remaining scattering channels. There
are no restrictions on how to choose the set ${\cal A}_1$.

It is useful to define the orthogonal projectors 
\beq
P^{(-)}_{{\cal A}_1} := \sum_{\alpha \in {\cal A}_1} P^{(-)}_{\alpha} \qquad \mbox{and} \qquad P^{(-)}_{{\cal A}_2} := \sum_{\alpha \in {\cal A}_2} P^{(-)}_{\alpha} ,
\label{hd:15}
\eeq
where by convention the one-body ($N$-body bound state) channels are
in ${\cal A}_2$.

If follows from  (\ref{hd:11}) that they satisfy
\beq
P^{(-)}_{{\cal A}_1} + P^{(-)}_{{\cal A}_2} = I .
\label{hd:16}
\eeq
This leads to the exact decomposition of the Hamiltonian given by
\[
H = P^{(-)}_{{\cal A}_1} H + P^{(-)}_{{\cal A}_2} H = \sum_{\alpha \in {\cal A}_1} P^{(-)}_{\alpha} H + \sum_{\alpha \in {\cal A}_2}P^{(-)}_{\alpha} H =
\]
\[
\sum_{\alpha \in {\cal A}_1} \left([P^{(-)}_{\alpha}H]_1 + \sum_{\{b \in {\cal P}_{N}' \vert b \supseteq a\}} {\cal C}_b (P^{(-)}_{\alpha})_b H_b\right)
\]
\beq
+\sum_{\alpha \in {\cal A}_2} \left([P^{(-)}_{\alpha} H]_1 + \sum_{\{b \in {\cal P}_{N}' \vert b \supseteq a\}} {\cal C}_b (P^{(-)}_{\alpha})_b H_b\right).
\label{hd:17}
\eeq
From (\ref{hd:14}), all of the completely connected parts in
(\ref{hd:17}) add up to $[H]_1$, and this vanishes if there are no
$N$-body interactions (note that the $N$-body bound state channels
only contribute to the completely connected parts of the
expression). It follows that
\beq
H = [H]_1 + \sum_{\alpha \in {\cal A}_1} \sum_{\{b \in {\cal P}_{N}' \vert b \supseteq a\}}{\cal C}_b (P^{(-)}_{\alpha})_b H_b + \sum_{\alpha \in {\cal A}_2}\sum_{\{b \in {\cal P}_{N}' \vert b \supseteq a\}}{\cal C}_b (P^{(-)}_{\alpha})_b H_b .
\label{hd:18}
\eeq
If there are no $N$-body interactions in the Hamiltonian, then the
contributions from the $N$-body bound states cancel with the
completely connected contributions from the scattering channels.

Channel truncated Hamiltonians are defined by 
\beq
H_{{\cal A}_1} := \sum_{\alpha \in {\cal A}_1} \sum_{\{b \in {\cal P}_{N}' \vert b \supseteq a\}} {\cal C}_b (P^{(-)}_{\alpha})_b H_b  
\label{hd:19}
\eeq
and
\beq
H_{{\cal A}_2} := [H]_1 + \sum_{\alpha \in {\cal A}_2} \sum_{\{b \in {\cal P}_{N}' \vert b \supseteq a\}} {\cal C}_b (P^{(-)}_{\alpha})_b H_b ,
\label{hd:20}
\eeq
where the completely connected part $[H]_1 $ (the $N$-body
interaction) is included in the set ${\cal A}_2$.  The individual
channel Hamiltonians are defined by
\beq
H_{\alpha} :=\sum_{\{b \in {\cal P}_{N}' \vert b \supseteq a\}} {\cal C}_b (P^{(-)}_{\alpha})_b H_b.  
\label{hd:20c}
\eeq
With this definition, equations (\ref{hd:19}) and (\ref{hd:20}) have the form

\beq
H_{{\cal A}_1} := \sum_{\alpha \in {\cal A}_1} H_{\alpha}
\label{hd:19a}
\eeq
and
\beq
H_{{\cal A}_2} := [H]_1 + \sum_{\alpha \in {\cal A}_2} H_{\alpha}.
\label{hd:20b}
\eeq
This decomposition has the feature that both
$H_{{\cal A}_1}$ and $H_{{\cal A}_2}$ 
are
expressed in terms of solutions of \textit{proper subsystem problems}
and a possible $N$-body interaction. $(P^{(-)}_{\alpha})_b$ is the
projection on the scattering states of $H_b$ if $H_b$ has scattering
states in the channel $\alpha$. These scattering states are
related to the exact channel $\alpha$ scattering states by
``turning off'' the interactions between particles in different
clusters of $b$ for $b \supseteq a$. It is always possible to add
additional $N$-body operators to (\ref{hd:19}) provided that they are
subtracted from (\ref{hd:20}).

Since both $H_b$ and $(P^{(-)}_{\alpha})_b$ are Hermitian and commute
with one another, it follows that both $H_{{\cal A}_1}$ and $H_{{\cal A}_2}$ 
are Hermitian. Also, since
\beq
\sum_{\alpha \in {\cal A}_1} \sum_{\{b \in {\cal P}_{N}' \vert b \supseteq a\}}{\cal C}_b (P^{(-)}_{\alpha})_b H_b = \sum_{\alpha \in {\cal A}_1} P^{(-)}_{\alpha} H - \sum_{\alpha \in {\cal A}_1} [P^{(-)}_{\alpha} H]_1 ,
\label{hd:21}
\eeq
$H_{{\cal A}_1}$ differs from the exact spectral projection of the
Hamiltonian, $P^{(-)}_{{\cal A}_1} H$, on the channels ${\cal A}_1$ by
the connected operator
\beq
W_I := [P^{(-)}_{{\cal A}_1} H]_1 = \sum_{\alpha \in {\cal A}_1} [P^{(-)}_{\alpha} H]_1  .
\label{hd:22}
\eeq
It follows from (\ref{hd:21}) that
\beq
P^{(-)}_{{\cal A}_1} H = H_{{\cal A}_1} + W_I .
\label{hd:23}
\eeq
The channel $\alpha$ scattering states of $H_{{\cal A}_1}$ are defined using wave operators given by
\beq
\Omega^{(-)}_{{\cal A}_1} (a) \Phi_{\alpha} = \lim_{t \to -\infty}e^{iH_{{\cal A}_1}t} e^{-iH_at}\Phi_{\alpha} .
\label{hd:24}
\eeq
Note that for the exact spectral projectors one has, for $\alpha \in {\cal A}_1$,
\[
\lim_{t \to -\infty} \Vert P^{(-)}_{{\cal A}_2} e^{i Ht }e^{-iH_at}\Phi_{\alpha}
\vert \phi_{o\alpha} \rangle \Vert = \Vert P^{(-)}_{{\cal A}_2}  P^{(-)}_{{\cal A}_1} \vert \phi_{o\alpha} \rangle \Vert = 0 ,
\]
since $P^{(-)}_{{\cal A}_2}$ and $P^{(-)}_{{\cal A}_1}$ project on
orthogonal subspaces.

Using (\ref{hd:24}), it follows that
\[
\Omega^{(-)}_{{\cal A}_1} (a) \Phi_{\alpha} =
\]
\[
\lim_{t \to -\infty} e^{iH_{{\cal A}_1}t}
\underbrace{e^{-iP^{(-)}_{{\cal A}_1}H t} e^{iP^{(-)}_{{\cal A}_1}H t}}_{I} e^{-iH_at}\Phi_{\alpha}
=
\]
\[
\lim_{t \to -\infty} e^{iH_{{\cal A}_1}t} e^{-iP^{(-)}_{{\cal A}_1}H t} \underbrace{( P^{(-)}_{{\cal A}_1}
+ P^{(-)}_{{\cal A}_2} )}_{I} \left(e^{iHt} e^{-iH_at}\Phi_{\alpha}\right) =
\]
\[
\lim_{t \to -\infty} e^{iH_{{\cal A}_1}t} e^{-iP^{(-)}_{{\cal A}_1}H t} P^{(-)}_{{\cal A}_1} \left(e^{iHt} e^{-iH_at}\Phi_{\alpha}\right) =
\]
\beq
\lim_{t \to -\infty} e^{iH_{{\cal A}_1}t} e^{-iP^{(-)}_{{\cal A}_1}H t} P^{(-)}_{{\cal A}_1} \left(\Omega^{(-)}(a) \Phi_{\alpha}\right) ,
\label{hd:25}
\eeq
where $P_{{\cal A}_1}^{(-)} \left(\Omega^{(-)}(a)
\Phi_{\alpha}\right) = \Omega^{(-)}(a)
\Phi_{\alpha}$ is the exact channel wave operator for the
$\alpha$ scattering states projected on the subspace of important
channels ${\cal A}_1$.
This shows, using the chain rule for wave
operators, that the channel $\alpha$ scattering states of $H_{{\cal A}_1}$
are related to the exact channel $\alpha$ scattering states
by the wave operator
\beq
\Omega_W^{(-)} := \lim_{t \to -\infty} e^{iH_{{\cal A}_1}t} e^{-iP^{(-)}_{{\cal A}_1}H t} = \lim_{t \to -\infty} e^{i\left(P^{(-)}_{{\cal A}_1}H - W_I\right) t} e^{-iP^{(-)}_{{\cal A}_1}H t} , 
\label{hd:26}
\eeq
which is a connected perturbation of the identity. It transforms
channel $\alpha$ eigenstates of the exact Hamiltonian (for states in
the set ${\cal A}_1$) to channel eigenstates of ${\cal H}_{{\cal
A}_1}$:
\beq
\Omega^{(-)}_{{\cal A}_1} (a) \Phi_{\alpha} = \Omega_{W}^{(-)} P_{{\cal A}_1}^{(-)} \left(\Omega^{(-)}(a) \Phi_{\alpha}\right) . 
\label{hd:27}
\eeq
Treating $W_I$ as a connected perturbation of $H_{{\cal A}_1}$ allows one to write the inverse relation given by
\beq
P_{{\cal A}_1}^{(-)} \left(\Omega^{(-)}(a) \Phi_{\alpha}\right) = \Omega_W^{(-) \dagger} \Omega^{(-)}_{{\cal A}_1} (a) \Phi_{\alpha} . 
\label{hd:28}
\eeq
It is important to note that although $\Omega_W^{(-)}$
depends on ${\cal A}_1$, it is independent of the specific channel
$\alpha \in {\cal A}_1$. While determining $W_I$ involves solving the
$N$-body problem, the observation that the scattering states of
$H_{{\cal A}_1}$ are related to the scattering states of the exact
Hamiltonian, projected onto the subspace of important channels,
implies that the spectral resolution of both operators are related by
$\Omega_W^{(-)}$.  This also means that the incoming scattering eigenstates of
$H$ and $H_{{\cal A}_1}$ in the channels ${{\cal A}_1}$ are identical up to
fully connected parts.  This does not require the full $N$-body solution.
This observation will be used in the next section to
show that $H_{{\cal A}_1}$ satisfies an optical theorem with the
channels $\alpha\in {\cal A}_1$.

The operator $\Omega_W^{(-)}$ also satisfies the intertwining relation:
\beq
e^{iH_{{\cal A}_1}s} \, \Omega_W^{(-)} = \Omega_W^{(-)} \, e^{iP^{(-)}_{{\cal A}_1}Hs} .
\label{hd:29}
\eeq
This follows from
\[
e^{iH_{{\cal A}_1}s} \, \Omega_W^{(-)} = \; e^{iH_{{\cal A}_1}s} \left( \lim_{t \to -\infty} e^{i\left(P^{(-)}_{{\cal A}_1}H - W_I\right) t} e^{-iP^{(-)}_{{\cal A}_1}H t} \right) =
\]
\[
e^{iH_{{\cal A}_1}s} \left( \lim_{t \to -\infty}  e^{i\left(P^{(-)}_{{\cal A}_1}H - W_I\right) t} e^{-iP^{(-)}_{{\cal A}_1}H t} \right) e^{-iP^{(-)}_{{\cal A}_1}Hs} e^{iP^{(-)}_{{\cal A}_1}Hs} =
\]
\[
\left( \lim_{(t + s) \to -\infty} e^{i\left(P^{(-)}_{{\cal A}_1}H - W_I\right)(s + t)} e^{-iP^{(-)}_{{\cal A}_1}H(t + s)} \right) e^{iP^{(-)}_{{\cal A}_1}Hs} = \, \Omega_W^{(-)} \, e^{iP^{(-)}_{{\cal A}_1}Hs} ,
\]
where the limit $t \rightarrow -\infty$ can be replaced with $(t + s)
\rightarrow -\infty$ because the limit is the same for any fixed $s$.

It follows from  the intertwining relation that 
\beq
\frac{1}{2\pi} \int ds \, e^{-isx} f(x) \, e^{isH_{{\cal A}_1}} \, \Omega_W^{(-)} =
\Omega_W^{(-)} \int ds \, e^{is P^{(-)}_{{\cal A}_1}H} \, \frac{1}{2\pi} e^{-isx} f(x) .
\label{hd:30}
\eeq
This means that
\beq
f(H_{{\cal A}_1}) \, \Omega_W^{(-)} = \Omega_W^{(-)} \, f(P^{(-)}_{{\cal A}_1}H).
\label{hd:31}
\eeq
For functions of the form $f(x) = \frac{1}{z- x}$, this gives
\beq
\left(E- H_{{\cal A}_1} + i\epsilon \right)^{-1} \Omega_W^{(-)} = \Omega_W^{(-)} \left(E- P^{(-)}_{{\cal A}_1}H + i\epsilon\right)^{-1} 
\label{hd:32}
\eeq
or
\beq
\left(E- H_{{\cal A}_1} + i\epsilon\right)^{-1} = \Omega_W^{(-)} \left(E- P^{(-)}_{{\cal A}_1}H + i\epsilon\right)^{-1} \Omega_W^{(-) \dagger} .  
\label{hd:33}
\eeq
In the next section, this will be used to prove the optical theorem.

\section{Optical Theorem}
\label{sec:4}

\noindent This section will show that the transition operator for
the truncated Hamiltonian, $H_{{\cal A}_1}$, satisfies an
optical theorem with the channels $\alpha\in {\cal A}_1$, which
shows that all of the scattered flux is in the channels ${\cal
  A}_1$.

The ${\cal A}_1$-transition operator for $2-2$ forward scattering in the two-body channel $\beta$, using (\ref{sc:5}) and (\ref{sc:14}), is  
\[
T^{\beta\beta}_{{\cal A}_1}(E_{\beta} + i\epsilon)  = 
\Phi_\beta^{\dagger} \, T_{{\cal A}_1}^{bb}(E_{\beta} + i\epsilon) \, \Phi_{\beta} =
\]
\beq
\Phi_\beta^{\dagger} \left(H_{{\cal A}_1}^b + H_{{\cal A}_1}^b \left(E_{\beta} - H_{{\cal A}_1} + i\epsilon\right)^{-1} H_{{\cal A}_1}^b\right) \Phi_\beta .
\label{ot:1}
\eeq
Taking the difference with $i \to -i$ leads to
\[
\Phi_\beta^{\dagger} \left(T_{{\cal A}_1}^{bb}(E_{\beta} + i\epsilon) - T_{{\cal A}_1}^{bb}(E_{\beta}-i\epsilon)\right) \Phi_\beta =
\]
\beq
\Phi_\beta^{\dagger} H_{{\cal A}_1}^b \left(\frac{-2 i\epsilon}{\left(E_{\beta} - H_{{\cal A}_1}\right)^2 + \epsilon^2}\right) H_{{\cal A}_1}^b \Phi_\beta .
\label{ot:2}
\eeq
Using (\ref{hd:33}), this becomes
\[
\Phi_\beta^{\dagger} \left(T_{{\cal A}_1}^{bb}(E_{\beta} + i\epsilon) - T_{{\cal A}_1}^{bb}(E_{\beta} - i\epsilon)\right) \Phi_\beta =
\]
\[
\Phi_\beta^{\dagger} H_{{\cal A}_1}^b \Omega_W^{(-)} \left(\frac{-2 i\epsilon}{\left(E_{\beta} - P^{(-)}_{{\cal A}_1}H\right)^2 + \epsilon^2}\right) \Omega_W^{(-) \dagger} H_{{\cal A}_1}^b \Phi_\beta ,
\]
and this means that
\[
\lim_{\epsilon \to 0} \Phi_\beta^{\dagger} \left(T_{{\cal A}_1}^{bb}(E_{\beta} + i\epsilon) - T_{{\cal A}_1}^{bb}(E_{\beta}-i\epsilon)\right) \Phi_\beta =
\]
\beq -2\pi i \, \Phi_\beta^{\dagger} H_{{\cal A}_1}^b \Omega_W^{(-)} \, \delta(P^{(-)}_{{\cal A}_1}H - E_{\beta}) \, \Omega_W^{(-) \dagger} H_{{\cal A}_1}^b \Phi_\beta .
\label{ot:3}
\eeq
The advantage of expressing this in terms of $P^{(-)}_{{\cal
    A}_1}H$ is that the completeness relation for the exact projected
Hamiltonian, which only involves states in the chosen set of important
channels ${\cal A}_1$, can be used to evaluate the delta function so
that (\ref{ot:3}) becomes
\beq
-2\pi i \sum_{\alpha \in {\cal A}_1} \Phi_\beta^{\dagger} H_{{\cal A}_1}^b \Omega_W^{(-)} \, \Omega^{(-)}(a) \Phi_{\alpha} \, \delta(E_{\alpha} - E_{\beta}) \, \Phi_{\alpha}^{\dagger} \,
\Omega^{(-) \dagger}(a) \, \Omega_W^{(-) \dagger} H_{{\cal A}_1}^b \Phi_\beta .
\label{ot:4}
\eeq
The exact channel $\alpha \in {\cal A}_1$ scattering states can be expressed in terms of the
channel $\alpha$ eigenstates of ${H}_{{\cal A}_1}$ using the relation in (\ref{hd:27}):
\beq
\Omega_W^{(-)} P^{(-)}_{{\cal A}_1} \left(\Omega^{(-)}(a) \Phi_{\alpha}\right) = \Omega_W^{(-)} \Omega^{(-)}(a) \Phi_{\alpha} = \Omega_{{\cal A}_1}^{(-)}(a) \Phi_{\alpha} ,
\label{ot:5}
\eeq
which means that (\ref{ot:4}) becomes
\beq
-2\pi i \sum_{\alpha \in {\cal A}_1} \Phi_\beta^{\dagger} H_{{\cal A}_1}^b \Omega_{{\cal A}_1}^{(-)}(a) \Phi_{\alpha} \, \delta (E_{\alpha} - E_{\beta}) \, \Phi_{\alpha}^{\dagger} \, \Omega_{{\cal A}_1}^{(-) \dagger}(a) H_{{\cal A}_1}^b \Phi_\beta .
\label{ot:6}
\eeq
Taking the imaginary part of both sides gives 
\[
2 Im\left\{\Phi_\beta^{\dagger} \, T_{{\cal A}_1}^{bb}(E_{\beta} + i\epsilon) \, \Phi_\beta\right\} =
\]
\[
-2\pi \sum_{\alpha \in {\cal A}_1} \Phi_\beta^{\dagger} H_{{\cal A}_1}^b \Omega_{{\cal A}_1}^{(-)}(a)
\Phi_{\alpha} \, \delta(E_{\alpha} - E_{\beta}) \, \Phi^{\dagger}_{\alpha} \, \Omega_{{\cal A}_1}^{(-) \dagger}(a) H_{{\cal A}_1}^b \Phi_\beta =
\]
\beq
-2\pi \sum_{\alpha \in {\cal A}_1} \int \vert \Phi_{\alpha}^{\dagger} \, T_{{\cal A}_1}^{ab}(E_{\beta} + i\epsilon) \, \Phi_{\beta} \vert^2 \, \delta(\sum_{i} E_{\alpha_i} - E_{\beta}) \, d\mathbf{p}_1 \cdots d\mathbf{p}_i .
\label{ot:7} 
\eeq
The right hand side of (\ref{ot:7}) is related to the total cross section
$\sigma_T$ by
\beq
\mbox{RHS} = - (2 \pi) \frac{v}{(2\pi)^4} \sigma_T ,
\label{ot:8}
\eeq
where $v$ is the relative velocity. This means that the total cross
section can be expressed as
\[
\sigma_T = - \frac{(2\pi)^3}{v} 2 Im\left\{\Phi_\beta^{\dagger} \, T_{{\cal A}_1}^{bb}(E_{\beta} + i\epsilon) \, \Phi_\beta\right\} =
\]
\beq
- \frac{2(2\pi)^3 \mu}{k}
\left (\frac{-1}{(2\pi)^2 \mu} \right ) Im\left\{F_{\beta\beta}\right\} = \frac{4\pi}{k} Im\left\{F_{\beta\beta}\right\} ,
\label{ot:9}
\eeq
where $F_{\beta\beta}:= - (2\pi)^2 \mu \, \Phi_\beta^{\dagger} \,
T_{{\cal A}_1}^{bb}(E_{\beta} + i\epsilon) \, \Phi_\beta$ is the
scattering amplitude for $2 \rightarrow 2$ forward scattering for the
two-body channel $\beta$, $k$ is the center-of-mass momentum, and
$\mu$ is the reduced mass. Equation (\ref{ot:9}) gives
the familiar form of the optical theorem:
\beq
\sigma_T = \frac{4\pi}{k} Im\left\{F_{\beta\beta}\right\} .
\label{ot:10}
\eeq
This means that all
of the scattered flux is in the channels ${\cal A}_1$. It is important
to note that, in this case, the discontinuity across the scattering
cuts in the resolvent does not receive contributions from
the
channels in ${\cal A}_2$. An identical analysis also applies to the Hamiltonian $H_{{\cal A}_2}$ by
interchanging the channels ${\cal A}_1$ with the channels
${\cal A}_2$.

\section{Benzce-Redish-Sloan Equations}
\label{sec:5}

\noindent While the properties of $H_{\cal A}$ and the associated scattering
theory were derived using time-dependent methods, computations
normally utilize time-independent methods. For many-body reactions,
differential cross sections are expressed in terms of transition
operators, $T^{ba}(z)$, which are operators on the $N$-particle
Hilbert space. They are 
related to the two-Hilbert space channel transition operators by 
\[
T^{\beta\alpha}_{{\cal A}_1}(E + i\epsilon) =
\Phi^{\dagger}_\beta H^b \Omega^{(-)}(a) \Phi_\alpha =
\Phi^{\dagger}_\beta T^{ba}(E + i \epsilon)  \Phi_\alpha ,  
\]
where $z = E + i\epsilon$ and $E$ is the \textit{on-shell} energy.
These are solutions of linear integral equations. By manipulating 
the equations so that they have a compact iterated kernel, which can be uniformly
approximated by a finite dimensional matrix, the solution involves
solving a large linear system. The equations derived by Bencze, Redish and
Sloan \cite{Sloan1972}\cite{Bencze1973}\cite{Redish1974} have this
property and are sufficiently flexible to be applicable to the
dynamical models governed by Hamiltonians of the form $H_{{\cal A}_1}$.  A
short derivation of these equations following \cite{Benoist} is given
below.

The transition operator for multichannel scattering
in the notation of this paper is
\beq
T^{ba}(z) = H^a + H^b G(z) H^a ,
\label{brs:1}
\eeq
where $G(z) = \left(z - H\right)^{-1}$ is the resolvent operator (or
Green's operator) and $z = E + i\epsilon$ is the complex energy.
If the
completely connected part of $H^b$ is zero (it doesn't contain any
$N$-body forces), then the operator decomposition of $H^b$ is
\beq
H^b = \sum_{c \in {\cal P}_{N}'} {\cal C}_c H_{c}^{b} .
\label{brs:2}
\eeq
This means that the transition operator can be expressed as
\beq
T^{ba}(z) = H^a + \sum_{c \in {\cal P}_{N}'} {\cal C}_c H_{c}^{b} G(z) H^a .
\label{brs:3}
\eeq
Using the second resolvent identity
\[
G(z) = G_c(z) + G_c(z) H^c G(z) ,
\]
the transition operator becomes
\[
T^{ba}(z) = H^a + \sum_{c \in {\cal P}_{N}'} {\cal C}_c H_{c}^{b} \left(G_c(z)+ G_c(z) H^c G(z)\right) H^a =
\]
\beq
H^a + \sum_{c \in {\cal P}_{N}'} {\cal C}_c H_{c}^{b} G_c(z) \left(H^a + H^c G(z) H^a\right) 
= H^a + \sum_{c \in {\cal P}_{N}'} {\cal C}_c H_{c}^{b} G_c(z) T^{ca}(z) ,
\label{brs:4}
\eeq
where $G_c(z) = \left(z - H_c\right)^{-1}$. Therefore, the transition operator $T^{ba}(z)$ satisfies
\beq
T^{ba}(z) = H^{a} + \sum_{c \in {\cal P}_{N}'} {\cal C}_c H^b_c G_c(z) T^{ca}(z) ,
\label{brs:5}
\eeq
and these are the equations derived by Bencze, Redish and Sloan.
These equations are coupled integral equations.
It follows from
(\ref{brs:5}) that the iterated kernel,
\beq
\sum_{c \in {\cal P}_{N}'} \sum_{d \in {\cal P}_{N}'} {\cal C}_c
{\cal C}_d H^b_c G_c(z)  H^c_d G_d(z), 
\label{brs:6}
\eeq
is connected since $\sum_{c \in {\cal P}_{N}'} {\cal C}_c H^b_c G_c(z)  H^c_d$ is connected or zero by (\ref{ce:23}).
Everything above holds if $H$ is replaced by  $H_{\cal A}$. 

\section{Identical Particles}
\label{sec:6}

\noindent For systems of identical nucleons, the Hilbert space is the
antisymmetrized subspace of the $N$-nucleon Hilbert space, and the
exchange symmetry can be used to reduce the number of coupled
scattering integral equations. In this section, the method in
\cite{BR-1978} is applied to treat integral equations of the form in
(\ref{brs:5}) involving identical particles.

Projectors on the symmetrized (antisymmetrized) subspace of the
Hilbert space are constructed using permutation operators.  ${\Pi}(N)$
is the group of permutations on $N$ objects. For a given permutation
$\sigma \in {\Pi}(N)$, $P_{\sigma}$ is the operator acting on a
$N$-particle basis vector $\vert \mathbf{k}_1,\mu_1, \cdots,
\mathbf{k}_{N},\mu_{N} \rangle$ defined by
\beq
P_{\sigma} \vert \mathbf{k}_1,\mu_1, \cdots, \mathbf{k}_{N},\mu_{N} \rangle =
(-)^{\vert \sigma \vert} \vert \mathbf{k}_{\sigma(1)},\mu_{\sigma (1)}, \cdots,\mathbf{k}_{\sigma(N)},\mu_{\sigma(N)}\rangle ,
\label{ip:1}
\eeq
where $\vert \sigma \vert = 0$ for integer spin particles and
even permutations of half-integer spin particles and $\vert \sigma
\vert = 1$ for odd permutations of half-integer spin particles. 

Permutations act on partitions by permuting the labels of particles in
each cluster.  The partition $b=\sigma (a)$ denotes the partition obtained from $a$
by using $\sigma$ to change the particle labels.  This is illustrated
by the following example
\beq
\sigma =
\left (
\begin{array}{ccccccc}
1&2&3&4&5&6&7\\
2&7&3&5&4&1&6\\  
\end{array}  
\right )
\label{ip:2}
\eeq
\beq
a=(124)(37)(65)
\label{ip:3}
\eeq
\beq
\sigma (a) = (275)(36)(14).
\label{ip:4}
\eeq
In this example, both partitions have one three-particle cluster and
two two-particle clusters.  Partitions related by permutations define
equivalence classes of partitions (permutation equivalent
partitions). The equivalence class containing partition $a$ is denoted
by $[a]$.  In the above example, this is the set of all distinct
seven-particle partitions with one three-particle cluster and two
two-particle clusters.  Any $b\in [a]$ can be expressed as $b= \sigma (a)$
for some $\sigma \in \Pi (N)$.

Permutations satisfying $\sigma (a) =a$ define a subgroup $\Pi_a(N)$
of $\Pi(N)$.  The elements of $\Pi_a(N)$ include permutations that
interchange particles in the same equivalence class of $a$ and
permutations that interchange different equivalence classes with the
same number of particles.  The subgroup ${\Pi}_a(N)$ has $N_a =
\prod_i n_{a_i}! \prod_j k_j!$ elements, where $n_{a_i}$ is the number
of particles in the $i^{th}$ cluster of $a$ and $k_j$ is the number of
clusters with $j$ particles.  In the above example, interchanging 3
with 7 and 6 with 5 leaves $a$ unchanged. Interchanging the (37) pair
with the (65) pair also leaves $a$ unchanged. The subgroup ${\Pi}_a(N)$
in the above example has $48=3!2!2!2!$ permutations.

For $\sigma' \in \Pi (N)$ satisfying  $\sigma' (a)= a'  \not =a$,
if $\sigma \in \Pi_a (N)$ then
$(\sigma' \sigma) (a)= \sigma'' (a) =  a'$.  This means that
$
(\sigma')^{-1}\sigma'' \in \Pi_a(N)$ or that $\sigma'$ and $\sigma''$ are
in the same left coset of
$\Pi_a(N)$.  It follows that each partition in $[a]$ can be
identified with a left coset of $\Pi_a(N)$.  The number of partitions
in $[a]$ is equal to the number of left cosets of $\Pi_a(N)$,
which by Lagrange's theorem is $N_{[a]}= \frac{N!}{N_a}$.  In the above example
there are $105=7!/48$ partitions in the class $[(275)(36)(14)]$.
As another example, the  equivalence class, $[(1)(2)(34)]$, contains
$\frac{4!}{2!2!}=3!$ partitions, and these include:
$(1)(2)(34)$, $(1)(3)(24)$, $(1)(4)(23)$, $(2)(3)(14)$, $(2)(4)(13)$, $(3)(4)(12)$.

Permutation operators, $P_\sigma$, are unitary since they are
products of unitary transposition operators.
The permutation operator $P_\sigma$ replaces the cluster translation operator in (\ref{ce:4}), that translates the clusters of 
the partition
$a$, by the cluster translation operator that translates the
clusters of the partition
$\sigma(a)$:
\beq
P_\sigma T_a (\mathbf{x}_1, \cdots ,\mathbf{x}_{n_a}) P_\sigma^{\dagger}
 =T_{\sigma (a)} (\mathbf{x}_1, \cdots , \mathbf{x}_{n_a}).
\label{ip:5}
\eeq
Since the operator
$O_a$ is defined using the cluster translation operator,
it follows from (\ref{ce:9}) and (\ref{ip:5}) that 
\beq
P_\sigma O_a P_\sigma^{\dagger} = P_\sigma O_a P_\sigma^{-1} = O_{\sigma (a)} .
\label{ip:6}
\eeq
Also, since $[O]_a$ and  $O^b_a$ are defined in terms of the $O_a$'s,
the transformation properties of these operators under
permutations is 
\beq
P_\sigma [O]_a P_\sigma^{-1} = [O]_{\sigma (a)}
\qquad \mbox{and} \qquad
P_\sigma O^b_a P_\sigma^{-1} = O^{\sigma (b)}_{\sigma(a)}. 
\label{ip:7}
\eeq

Symmetrizers (antisymmetrizers)
are defined by
\beq
R := \frac{1}{N!} \sum_{\sigma \in \Pi(N)} P_{\sigma} .
\label{ip:8}
\eeq
The symmetrizer (antisymmetrizer) $R$ is an orthogonal projection
operator on the $N$-particle Hilbert space that satisfies
\beq
R = R^2 = R^{\dagger} \qquad \mbox{and} \qquad R = P_{\sigma} R = R P_{\sigma} \qquad \forall \; \sigma \in \Pi (N) .
\label{ip:9}
\eeq
The first relation in (\ref{ip:9}) can be obtained using
\[
R^2 = \frac{1}{N!} \, \frac{1}{N!} \sum_{\sigma \in \Pi(N)} \left(\sum_{\sigma' \in {\Pi}(N)} P_{\sigma} P_{\sigma'}\right) = \frac{1}{N!} \, \frac{1}{N!} \underbrace{\sum_{\sigma \in \Pi (N)}}_{= N!} \underbrace{\left(\sum_{\sigma'' \in \Pi(N)} P_{\sigma''}\right)}_{= N! \, R} =
\]
\beq
R = R^{\dagger},
\label{ip:10}
\eeq
where $\sigma'' = \sigma \sigma'$ for a fixed $\sigma$.  The adjoint
replaces $P_{\sigma}$ by $P_{\sigma^{-1}}$, which results in
the same operator when summed over all permutations.  The second
relation in (\ref{ip:9}) can be obtained from the definitions of $R$
and $P_{\sigma}$, which simply relabels the permutations in the sum
of (\ref{ip:8}).

Symmetrizers (antisymmetrizers) constructed from permutations in
the subgroup that leaves the partition $a$ unchanged are defined by
\beq
R_a := \frac{1}{N_a} \sum_{\sigma \in {\Pi}_a(N)} P_{\sigma} .
\label{ip:11}
\eeq
These operators symmetrize the particles in each cluster of $a$ and
identical clusters of $a$.  Therefore, for any partition $a$,
the symmetrizer (antisymmetrizer) in (\ref{ip:8}) can be expressed as

\beq
R = \frac{1}{N_{[a]}} \sum_{a'\in [a]} P_{a'a} R_a = \frac{1}{N_{[a]}} \sum_{a'\in [a]} R_a P_{aa'} ,
\label{ip:12}
\eeq
where $P_{aa'}$ is any permutation operator that transforms $a$ to $a'$ and $N_{[a]}$
is the number of partitions in the equivalence class $[a]$.   
The sums are over all partitions that can be obtained
from $a$ by permutation. 

The Hilbert space for a system of $N$ identical particles is
the range of symmetrizer (antisymmetrizer), $R$, on the $N$-particle Hilbert space. The initial
and final scattering states need to be projected on this subspace so
that the unit normalized states are
\beq
\vert \psi \rangle \rightarrow \vert \psi \rangle_R :=
\frac{R \vert \psi \rangle}{\langle \psi \vert R^{\dagger} R \vert \psi \rangle^{1/2}} =
\frac{R \vert \psi \rangle}{\langle \psi \vert R \vert \psi \rangle^{1/2}} .
\label{ip:13}
\eeq
For states $\vert \psi \rangle$ satisfying $R_a \vert \Psi \rangle = \vert \psi \rangle$ and $\langle \psi \vert P_{\sigma} \psi \rangle =0$ for
$\sigma (a) \not=a$, the normalization coefficient is
%
\beq
\langle \psi \vert R \vert \psi \rangle^{-\frac{1}{2}} = \sqrt{N_{[a]}}.
\label{ip:13}
\eeq
For a wave function $\vert \psi \rangle$, with $N_{[a]}$ partition equivalent
channels, the normalized
wave function has the familiar form
\[
\vert \psi_{[a_o]} \rangle = 
\]
\[
\sqrt{N_{[a]}}\sum_{a \in [a_o]}\frac{1}{N_{[a]}}P_{aa_o}\vert \psi_{a_o} \rangle =
\sum_{a \in [a_o]}\frac{1}{\sqrt{N_{[a]}}}P_{aa_o}\vert \psi_{a_o} \rangle .
\]
Unit normalized wave functions are needed for matrix elements of the
(unitary) scattering operator to be interpreted as a probability
amplitude.

For distinguishable particles, if $\vert \alpha \rangle$ and $\vert
\beta \rangle$ are sharp-momentum initial and final channel states
with energy $E$, then
\beq
H^a \vert \alpha \rangle = \left(H^a + H_a - E\right) \vert \alpha \rangle = \left(H - E\right) \vert \alpha \rangle
\label{ip:14}
\eeq
\beq
\langle \beta \vert H^b = \langle \beta \vert \left(H^b + H_b - E\right) = \langle \beta \vert \left(H - E\right) ,
\label{ip:15}
\eeq
where $(H_a-E)\vert \alpha \rangle= (H_b-E)\vert \beta \rangle=0$. 
On-shell sharp-momentum transition matrix elements can
be expressed as
\beq
\begin{aligned}
&\langle \beta \vert T^{ba}(E + i\epsilon) \vert \alpha \rangle =
\langle \beta \vert \left(H^a + H^b \frac{1}{E - H + i\epsilon} H^a\right) \vert \alpha \rangle = \\& \langle \beta \vert \left ((H - E) + (H - E) \frac{1}{E - H + i\epsilon} (H - E)\right) \vert \alpha \rangle = \\&\langle \beta \vert \tilde{T}(E + i\epsilon) \vert \alpha \rangle ,
\end{aligned}
\label{ip:16}
\eeq
where $\tilde{T}(E + i\epsilon)$ is defined as
\beq
\tilde{T}(E + i\epsilon) := \left(H - E\right) + \left(H - E\right) \frac{1}{E - H + i\epsilon} \left(H - E\right).
\label{ip:17}
\eeq
If the Hamiltonian satisfies $[H,R]=0$, then $[\tilde{T}(E + i\epsilon),R]=0$ 
because $\tilde{T}(E + i\epsilon)$
is a function of $H$.  It follows that 
\[
\langle \beta \vert R \tilde{T}(E + i\epsilon) R \vert \alpha \rangle = 
\langle \beta \vert \tilde{T}(E + i\epsilon) R \vert \alpha \rangle =
\]
\beq
\langle \beta \vert R \tilde{T}(E + i\epsilon) \vert \alpha \rangle = 
\langle \beta \vert R \tilde{T}(E + i\epsilon) P_{\sigma}\vert \alpha \rangle ,
\label{ip:18}
\eeq
where $P_{\sigma}$ is an arbitrary permutation operator.  In this case, the
properly normalized on-shell transition operator from $\vert \alpha_o \rangle$ to
$\vert \beta_o \rangle$ is
\[
\frac{1}{N_{[b_o]}}
\frac{
\sum_{b \in [b_o]}\langle \beta_o \vert R_{b_o}
P_{b_ob} \tilde{T}(E+i\epsilon)P_{\sigma} \vert \alpha_o \rangle
}
{
\langle \alpha_o \vert R \vert \alpha_o \rangle^{1/2}
\langle \beta_o \vert R \vert \beta_o \rangle^{1/2}
} =
\]
\beq
\frac{1}{N_{[b_o]}}
\frac{
\sum_{b \in [b_o]}\langle \beta_o \vert R_{b_o}
P_{b_ob} \tilde{T}(E+i\epsilon) \vert \alpha \rangle
}
{
\langle \alpha_o \vert R \vert \alpha_o \rangle^{1/2}
\langle \beta_o \vert R \vert \beta_o \rangle^{1/2} 
} .
\label{ip:19}
\eeq
In this expression, the matrix element is independent of the choice
of $\vert \alpha \rangle = P_{\sigma}\vert \alpha_o \rangle$.

The normalization condition in (\ref{ip:13}) gives  
\beq
\sqrt{\frac{N_{[a_o]}}{N_{[b_o]}}}
\sum_{b \in [b_o]}\langle \beta_o \vert R_{b_o}
P_{b_o} \tilde{T}(E+i\epsilon) \vert \alpha \rangle .
\label{ip:20}
\eeq 
This can be expressed in terms of the operator $T^{ba}$ by
\beq
\sqrt{\frac{N_{[a_o]}}{N_{[b_o]}}}
\sum_{b \in [b_o]}\langle \beta_o \vert R_{b_o}
P_{b_ob} {T}^{ba} (E+i\epsilon) \vert \alpha \rangle .
\label{ip:21}
\eeq
This expression is only valid on shell, but it is possible to construct
equations for the operator
\beq
T^{[b_o]a}=
\sum_{b \in [b_o]}R_{b_o}
P_{b_ob} {T}^{ba} (E+i\epsilon) .
\label{ip:22}
\eeq
After solving for $T^{[b_o]a}$, it can be evaluated on shell
and multiplied by the numerical factors in (\ref{ip:21}).

To construct integral equations for $T^{[b_o]a}$, one uses the integral equations
in (\ref{brs:5})
for $T^{ba} (E+i\epsilon)$ in (\ref{ip:22}) to get
\beq
T^{[b_o]a}=
\sum_{b \in [b_o]}R_{b_o}
P_{b_o b} \left ( H^a + \sum_{c\in \cal{P}'_N}{\cal C}_c H^{b}_{c} G_c(E+i\epsilon)
{T}^{ca} (E+i\epsilon ) \right ).
\label{ip:23}
\eeq
To get an integral equation for $T^{[b_o]a}$, the second term on the
right-hand side of (\ref{ip:23}) needs to be expressed in terms of $T^{[c_o]a}$. 

The first step is to choose a permutation $\sigma$ satisfying
$\sigma (c_o)=c$, where $c_o \in[c]$ is an arbitrary but fixed
element of $[c]$.  For this permutation, let 
$P_{c c_o}=P_{\sigma}$.  It follows from (\ref{ip:6}) and (\ref{ip:7})  that
\beq
H^{b}_{c} G_c(E+i\epsilon ) = P_{cc_o}
H^{\sigma^{-1} (b)}_{c_o} G_{c_o}(E+i\epsilon )P_{c_oc}, 
\label{ip:24}
\eeq
where $P_{c_oc} = P^{-1}_{cc_o}$.

With this substitution the second term on the right side of (\ref{ip:23})
becomes
\[
\sum_{b \in [b_o]}R_{b_o}
P_{b_ob} \sum_{c\in \cal{P}'_N}{\cal C}_c H^{b}_{c} G_c(E+i\epsilon)
{T}^{ca} (E+i\epsilon ) =
\]
\beq
\sum_{b \in [b_o]}R_{b_o}
P_{b_ob} \sum_{c\in \cal{P}'_N}{\cal C}_c
P_{c c_o} H^{\sigma^{-1} (b)}_{c_o} G_{c_o}(E+i\epsilon) P_{c c_o}^{-1}
{T}^{ca} (E+i\epsilon ) .
\label{ip:25}
\eeq
The sum over partitions can be replaced by a sum over equivalence classes of
partitions and a sum over partition equivalent partitions in a given
equivalence class.  With
this decomposition, ${\cal C}_c= {\cal C}_{[c]}$ and $c_o$
is the same for all permutation equivalent partitions.

Since
$P_{b_ob}= P_{\sigma'}$ for some $\sigma'$, it follows that   
$P_{b_ob}P_{cc_o}=  P_{\sigma'\sigma} = P_{\sigma''}$.
One can then define $P_{b_o \sigma^{-1}(b)}:= P_{\sigma''} $.
With this definition, (\ref{ip:25}) can be written as
\beq
\sum_{[c_o]\in \cal{P}'_N}{\cal C}_{[c_o]} \sum_{c \in [c_o]}
\sum_{b \in [b_o]}R_{b_o}
P_{b_o\sigma^{-1}(b)} 
H^{\sigma^{-1}(b)}_{c_o} G_{c_o}(E+i\epsilon) P_{c c_o}^{-1}
{T}^{ca} (E+i\epsilon).
\label{ip:26}
\eeq
Since for each $c\in[c_o]$, $\sigma$ is a fixed permutation,
summing over all $b\in[b_o]$ is the same as summing over all
$\sigma^{-1} (b)\in [b_o]$.  This is independent of the specific $c\in[c_o]$.
Letting $b'=\sigma^{-1} (b)$, (\ref{ip:26}) becomes
\beq
\sum_{[c_o]\in \cal{P}'_N}{\cal C}_{[c_o]}
\sum_{b' \in [b_o]}R_{b_o}
P_{b_o b'} 
H^{b'}_{c_o} G_{c_o}(E+i\epsilon)  \sum_{c \in [c_o]} P_{c_o c}
{T}^{ca} (E+i\epsilon).
\label{ip:26}
\eeq
Since the choice of $P_{c_oc}$ was arbitrary, it could be replaced by
$P_{\sigma}P_{c_oc}$ for any $\sigma\in \Pi_{c_o}(N)$.  Averaging over all of the permutations in $\Pi_{c_o}(N)$ gives
\beq
\sum_{[c_o]\in \cal{P}'_N}{\cal C}_{[c_o]}
\sum_{b' \in [b_o]}R_{b_o}
P_{b_o b'} 
H^{b'}_{c_o} G_{c_o}(E+i\epsilon)  \sum_{c \in [c_o]} R_{c_o} P_{c_o c}
{T}^{ca} (E+i\epsilon), 
\label{ip:27}
\eeq
which can be expressed in terms of  ${T}^{[c_o]a} (E+i\epsilon)$ as
\beq
\sum_{[c_o]\in \cal{P}_N'}{\cal C}_{[c_o]}
\sum_{b' \in [b_o]}R_{b_o}
P_{b_o b'} 
H^{b'}_{c_o} G_{c_o}(E+i\epsilon)
{T}^{[c_o]a} (E+i\epsilon).  
\label{ip:28}
\eeq
In this equation, the number of coupled equations is equal to the
number of equivalence classes of partitions rather
than the number of partitions.
Combining (\ref{ip:22}), (\ref{ip:23}) and  (\ref{ip:27}) gives the following integral equation for
${T}^{[b_o]a} (E+i\epsilon)$:
\[
{T}^{[b_o]a} (E+i\epsilon) = \sum_{b' \in [b_o]}R_{b_o}
P_{b_o b'}H^{a}
\]
\beq
+
\sum_{[c_o]\in \cal{P}'_N}{\cal C}_{[c]}
\sum_{b' \in [b_o]}R_{b_o}
P_{b_o b'} 
H^{b}_{c_o} G_{c_o}(E+i\epsilon)
{T}^{[c_o]a} (E+i\epsilon ).  
\label{ip:29}
\eeq
After solving the system for $T^{[b_o]a_{o}}(E + i\epsilon)$, if it is
evaluated between internally symmetrized channel states, $\vert \alpha_o
\rangle$ and $\vert \beta_o \rangle$, then the matrix elements need to be
multiplied by $\sqrt{\frac{N_{[a_o]}}{N_{[b_o]}}}$ to get the correct
normalization. Since the normalization factors only depend on the
partitions, they can be absorbed into the equations by replacing
$T^{[b_o]a_{o}}(E + i\epsilon)$ by the symmetrized transition operator
defined by
\beq
T^{[b_o]a_o}_{sym} (E + i\epsilon) :=
\sqrt{\frac{N_{[a_o]}}{N_{[b_o]}}} T^{[b_o]a_{o}}(E + i\epsilon) .
\label{ip:30}
\eeq
The equations in (\ref{ip:29}) are for a general
permutation-symmetric $N$-body Hamiltonian, and they are also valid for
the truncated Hamiltonian, $H_{{\cal A}_1}$, provided ${\cal A}_1$
contains all channels related by permutations.  The number of equations
for truncated Hamiltonians depend on the choice of retained
channels ${\cal A}_1$.

The differential cross section has the form
\beq
d\sigma = \frac{(2\pi)^4}{v} \frac{N_{[a_o]}}{N_{[b_o]}}
\vert \langle \beta_o' \vert T^{[b_o]a_o} (E+i \epsilon ) \vert
\alpha_o \rangle \vert^2 \delta (E'_b-E_a)\delta (\mathbf{P}' - \mathbf{P})
\prod d\mathbf{p}_i
\eeq
when it is expressed in terms of the operator $T^{[b_o]a_o}$.

\section{The Structure of the Truncated Equations}
\label{sec:7}

\noindent The dynamical equations in (\ref{ip:29}) are abstract. This section considers the
structure of the simplest approximation, where only 2-cluster
channels are retained, in more detail.

In general, the truncated Hamiltonian in (\ref{hd:19}) has the form
\[
H_{{\cal A}_1} = \sum_a {\cal C}_a (H_{{\cal A}_1})_a .
\]
When ${\cal A}_1$ only includes 2-cluster channels, 
$(H_{{\cal A}_1})_a=0$ unless $a$ is a 2-cluster partition.
For 2-cluster partitions, ${\cal C}_a=1$ and, given
partitions $a$ and $b$, 
\[
(H_{{\cal A}_1})^b_a = (H_{{\cal A}_1})_a - (H_{{\cal A}_1})_{a\cap b}.
\]
If $a \not= b$, then ${a\cap b}$ will have more than two clusters.
This implies that
\[
(H_{{\cal A}_1})_{a\cap b}=0 \qquad \mbox{when} \qquad a\not=b .
\]
It follows that
\[
(H_{{\cal A}_1})^b_a = (H_{{\cal A}_1})_a \bar{\delta}_{ab} =
[H_{{\cal A}_1}]_a \bar{\delta}_{ab} ,
\]
where 
\[
\bar{\delta}_{ab} = 1 -{\delta}_{ab}.
\]
The channel $\alpha$ eigenstates of $(H_{{\cal A}_1})_a$
have the form
\[
\vert \alpha \rangle := 
\vert \alpha_1, s_1, \mu_1, \mathbf{p}_1 \rangle 
\otimes \vert \alpha_2, s_2, \mu_2, \mathbf{p}_2 \rangle ,
\]
where the $\alpha_i$ labels the bound subsystems. The energy
of this state is
\[
E_{\alpha_a} =
\frac{\mathbf{p}^2_1}{2 M_1}- e_1+
\frac{\mathbf{p}^2_2}{2 M_2}- e_2 ,
\]
and this is the sum of the kinetic energy minus the binding energy of each bound cluster ($M_i$ is the total mass of each bound cluster).

The terms in the kernel and driving term of (\ref{ip:29}) become 
\beq
(H_{{\cal A}_1})_{c}^{b} = \bar{\delta}_{bc}
[H_{{\cal A}_1}]_{c} = \bar{\delta}_{bc}
\int \vert \gamma_{c} \rangle E_{\gamma_{c}} d\gamma_{c} \langle \gamma_{c} \vert
\label{teq:1}
\eeq
and
\beq
(H_{{\cal A}_1})_{c}^{b} (G_{{\cal A}_1})_c(E + i\epsilon) =
\bar{\delta}_{bc} \int \vert \gamma_{c} \rangle \frac{E_{\gamma_{c}}
d\gamma_{c}}{E - E_{\gamma_{c}} + i\epsilon} \langle \gamma_{c} \vert 
\label{teq:2} ,
\eeq
where
\[
d\gamma_{c} = d\mathbf{p}_{c1} d\mathbf{p}_{c2}.
\]
In this case, the symmetrized equations in (\ref{ip:29}), with ${\cal C}_c = 1$ for 2-cluster partitions, are
\beq
\begin{aligned}
\langle \beta_{o} \vert & T^{[b_o]a_o}(E + i\epsilon) \vert \alpha_{o} \rangle
= \sum_{b \in [b_o]} \sum_{c\not=b} \int \langle \beta_{o}
\vert R_{b_o}P_{b_{o}b} \vert \gamma_{c} \rangle E_{\gamma_c}
d\gamma_c \langle \gamma_{c} \vert \alpha_{o} \rangle \\ & +
\sum_{b \in [b_o]} \sum_{c_o \not= b} \sum_{[c]} \int
\langle \beta_{o}\vert R_{b_o} P_{b_{o}b} \vert \gamma_{c_o} \rangle
\frac{E_{\gamma_{c_o}} d\gamma_{c_o}}{E - E_{\gamma_{c_o}} + i\epsilon}
\langle \gamma_{c_o} \vert T^{[c]a_o}(E + i\epsilon) \vert \alpha_{o} \rangle .
\end{aligned}
\label{teq:3}
\eeq
If $\vert \beta_o \rangle$ is internally symmetrized,
$\vert \beta_o \rangle = R_{\beta_p}\vert \beta_o \rangle$,  
then (\ref{teq:3}) becomes
\beq
\begin{aligned}
\langle \beta_{o} \vert & T^{[b_o]a_o}(E + i\epsilon) \vert \alpha_{o} \rangle
= \sum_{b \in [b_o]} \sum_{c\not=b} \int \langle \beta 
\vert \gamma_{c} \rangle E_{\gamma_c}
d\gamma_c \langle \gamma_{c} \vert \alpha_{o} \rangle \\ & +
\sum_{b \in [b_o]} \sum_{c_o \not= b} \sum_{[c_o]} \int
\langle \beta \vert \gamma_{c_o} \rangle
\frac{E_{\gamma_{c_o}} d\gamma_{c_o}}{E - E_{\gamma_{c_o}} + i\epsilon}
\langle \gamma_{c_o} \vert T^{[c_o]a_o}(E + i\epsilon) \vert \alpha_{o} \rangle .
\end{aligned}
\label{teq:3b}
\eeq
These solutions need to be multiplied by
$\sqrt{\frac{N_{[a_o]}}{N_{[b_o]}}}$ in order to get the properly normalized
transition matrix elements.

What is needed as input are the overlap matrix elements
\[
\langle \beta_{o} \vert R_{b_o} P_{b_{o}b} \vert \gamma_{c_o} \rangle 
\]
for all $b \in [b_o]$.  Each one has an overall momentum conserving delta
function with a rotationally covariant kernel that depends on one initial
and one final relative momentum variable.  They have the general structure
\[
\int \sum_{n_b} \langle \tilde{\mathbf{p}}_{b_o}, n_{b_o}, \beta_o
\vert R_{b_o} P_{b_{o}b} \vert 
\tilde{\mathbf{p}}_{b}, n_{b}, \beta \rangle d\tilde{\mathbf{p}}_b
\langle \tilde{\mathbf{p}}_{b}, n_{b}, \beta  
\vert \tilde{\mathbf{p}}_{c_o}, n_{c_o},
\gamma_{c_o} \rangle ,
\]
where the $\tilde{\mathbf{p}}_a$ are the relative momenta between the clusters of $a$.
For each fixed $\tilde{\mathbf{p}}_b$, the remaining independent variables are integrated out.

\section{Bound States}
\label{sec:8}

\noindent While the main goal of this work is to investigate
the role of
different scattering channels on many-body reactions,
the result is a truncated Hamiltonian
that may also have bound states. If the original Hamiltonian has no
$N$-body interactions, then the projection of the exact Hamiltonian on
the bound states must exactly cancel with the connected part of the
projection of the Hamiltonian on the scattering states (using the
equation in (\ref{hd:17})). The dominant part of the projection of the
Hamiltonian on the bound states may be due to $N$-body contributions
from a limited number of important incoming or outgoing scattering
channels.  If these limited scattering channels are major contributors
to the connected part of the exact scattering states, then they should
provide major contributions to the bound states of the system.

The decomposition in (\ref{hd:19}) can be used to determine which scattering
channels are most responsible for the binding. An interesting example
is the system consisting of two protons and four neutrons. This system has
a bound state, $^6{He}$, which is a halo nucleus. The expectation is that
this system can be modeled as an alpha particle interacting with two loosely bound halo neutrons. The six-body bound state arises from the connected part of the
complete set of scattering states. An interesting question is how
much of the binding is due to the subset of $\alpha-n-n$ scattering channels.
This can be investigated by searching for bound states of the truncated
Hamiltonian $H_{{\cal A}_1}$, where ${\cal A}_1$ consists of all of
the $\alpha-n-n$ scattering channels. The interesting thing about this system is that there are no bound states consisting of two protons and three neutrons, $^5{He}$, and there are no bound states consisting of two neutrons. The interactions for this reaction mechanism are constructed from the $\alpha-n$ and
$n-n$ scattering states where the third ``particle'' acts as a
non-interacting spectator.

In this example, there are six equivalent scattering channels that differ by which pair
of neutrons are bound in the alpha particle.  These channels can be
labeled by the pair of neutrons in the alpha particle:
\[
\alpha_{ij},  a_{\alpha_{ij}} =(p_1 p_2 n_i n_j)(n_k)(n_l) 
\]
where $p$ and $n$ are the protons and neutrons, respectively. Here $i$ and $j$ label the neutrons in the alpha particle and
$k$ and $l$ label the asymptotically free neutrons.
There are 6 permutation equivalent channels and partitions:
$\alpha_{12},\alpha_{13},\alpha_{14},\alpha_{23},\alpha_{24},\alpha_{34}$.
The partitions that appear in the Hamiltonian
\[
H_{{\cal A}_1} = \sum_{a \in {\cal P}_{N}'} {\cal C}_a (H_{{\cal A}_1})_a
\]
are the 3-cluster partitions, $(p_1 p_2 n_i n_j)(n_k)(n_l)$,
and the 2-cluster partitions that include
\[
(p_1 p_2 n_i n_j n_k)(n_l),
(p_1 p_2 n_i n_j n_l)(n_k)
\, \mbox{and}\, 
(p_1 p_2 n_i n_j)(n_k n_l) ,
\]
where there are six combinations of $ij$.

The following short-hand notation is used in what follows:
\[
\vert \alpha_{ij} \rangle =
\vert \alpha, \mathbf{p}_{ij} \rangle
\otimes \vert \mathbf{p}_k, \mu_k \rangle \otimes \vert \mathbf{p}_l,\mu_l \rangle
\]
\[
\vert \alpha_{(ij)(k)^-} \rangle =
\vert (\alpha, \mathbf{p}_{ij}, \mathbf{p}_k, \mu_k)^- \rangle
\otimes \vert \mathbf{p}_l, \mu_l \rangle
\]
\[
\vert \alpha_{(ij)(l)^-} \rangle =
\vert (\alpha, \mathbf{p}_{ij}, \mathbf{p}_l, \mu_l)^- \rangle
\otimes \vert \mathbf{p}_k, \mu_k \rangle
\]
\[
\vert \alpha_{(ij)^-} \rangle =
\vert \alpha, \mathbf{p}_{ij} \rangle \otimes \vert
(\mathbf{p}_k, \mu_k, \mathbf{p}_l, \mu_l)^- \rangle
\]
\[
E_{ij} = \frac{\mathbf{p_{ij}^2}}{8m_N} + \frac{\mathbf{p}_k^2}{2m_k} +
\frac{\mathbf{p}_l^2}{2m_l} - e_{\alpha}
\]
\[
d\alpha_{ij} = d\mathbf{p}_{ij} d\mathbf{p}_{k} d\mathbf{p}_{l}
\]
where $\mathbf{p}_{ij}= \mathbf{p}_i + \mathbf{p}_j$ and $e_{\alpha}$
is the binding energy of the $\alpha$ particle.
The kinetic energy and interaction terms are defined by
\[
K_{ij} = \int \sum_{\mu_l,\mu_k}
\vert \alpha_{ij} \rangle d\alpha_{ij} E_{ij} \langle \alpha_{ij} \vert  
\]
\[
H_{ij,k} =
\int \sum_{\mu_l,\mu_k} \left (
\vert \alpha_{(ij)(k)^-} \rangle 
d\alpha_{ij} E_{ij}
\langle \alpha_{(ij)(k)^-} \vert -
\vert \alpha_{(ij)} \rangle 
d\alpha_{ij} E_{ij}
\langle \alpha_{(ij)} \vert \right )
\]
\[
H_{ij,l} =
\int  \sum_{\mu_l,\mu_k} \left ( 
\vert \alpha_{(ij)(l)^-} \rangle 
d\alpha_{ij} E_{ij}
\langle \alpha_{(ij)(l)^-} \vert -
\vert \alpha_{(ij)} \rangle 
d\alpha_{ij} E_{ij}
\langle \alpha_{(ij)} \vert \right ) 
\]
\[
H_{kl} =
\int \sum_{\mu_l,\mu_k}\left ( 
\vert \alpha_{(ij)^-} \rangle 
d\alpha_{ij} E_{ij}
\langle \alpha_{(ij)^-} \vert -
\vert \alpha_{(ij)} \rangle 
d\alpha_{ij} E_{ij}
\langle \alpha_{(ij)} \vert \right ) .
\]
Using this notation, the Hamiltonian for the $\alpha-n-n$ channels is
\[
H_{{\cal A}_1}= \sum_{ij}
(K_{ij} + H_{ij,k} + H_{ij,l} + H_{kl}) ,
\]
where the sum is over all six pairs of $ij$ corresponding to 24 partitions.
The input are $n-n$ and $n-\alpha $
scattering states.  Mathematically, this is a coupled three-body system.
This Hamiltonian can be diagonalized to determine if
this reaction mechanism is sufficiently rich to support a bound state.

Using
\[
\left(E- (H_{{\cal A}_1})_a\right) \vert \Psi \rangle = (H_{{\cal A}_1})^a  \vert \Psi \rangle
\]
and
\[
\sum_{a \in {\cal P}_{N}'} {\cal C}_a = 1 , 
\]
it follows that the six-body bound state is a solution of the generalized
eigenvalue problem
\[
\vert \Psi \rangle = \sum_{a} {\cal C}_a (E- (H_{{\cal A}_1})_a)^{-1}
(H_{{\cal A}_1})^a \vert \Psi \rangle ,
\]
where the right-hand side of this equation is connected by (\ref{ce:23}).
The sum only involves 2-cluster and 3-cluster partitions.
Note that this particular form of the equation is known to have spurious
solutions \cite{Glockle}, so any solutions need to be checked to make sure
that they also satisfy the Schr\"odinger equation.

\section{The Relativistic Case}
\label{sec:9}

\noindent The same analysis can be applied to a relativistically invariant
quantum theory, with some non-trivial differences.  Relativistic
invariance in a quantum theory requires that quantum observables
cannot be used to distinguish inertial coordinate systems.  This
implies that the dynamics of the system is given by a unitary
representation, $U(\Lambda ,a)$, of the Poincar\'e group
\cite{wigner:1939cj} (semi-direct product of the Lorentz group ($\Lambda$) and
spacetime translation group ($a$)).  Unitary transformations preserve
the quantum observables:
quantum probabilities, expectation values and ensemble averages.

The Poincar\'e group is a ten parameter group (three translations,
three rotations, three rotationless boosts and time translation).  The
infinitesimal generators of $U(\Lambda ,a)$ are the Hamiltonian, $H$,
the linear momentum operator, $\mathbf{P}$, the angular momentum
operator, $\mathbf{J}$, and the rotationless boost generator
$\mathbf{K}$.
These are self-adjoint operators. They satisfy the Poincar\'e
commutation relations:
\beq
[P^{\mu},P^{\nu}]=0 \qquad [J^i, P^j]= i \epsilon^{ijk}P^k
\qquad [J^i, J^j]= i \epsilon^{ijk}J^k
\label{sr:1}
\eeq
\beq
[J^i, K^j]= i \epsilon^{ijk}K^k \qquad
[K^i, K^j]= -i \epsilon^{ijk}J^k
\label{sr:2}
\eeq
\beq
[K^i,P^i] = i\delta^{ij} H
\qquad
[K^i,H] = i P^i.
\label{sq:3}
\eeq
The relativistic analog of diagonalizing the Hamiltonian is to
decompose $U(\Lambda,a)$ into a direct integral of irreducible
representations. This is equivalent to simultaneously diagonalizing the
invariant mass and the spin Casimir operators of the Lie algebra defined by 
\beq
M^2= H^2 - \mathbf{P}^2 \qquad \mbox{and} \qquad \mathbf{S}^2 = W^2 /M^2 ,
\label{sr:4}
\eeq
where $W^{\mu}$ is the Pauli-Lubanski vector
\beq
W^{\mu}=(\mathbf{P}\cdot \mathbf{J}, H\mathbf{J}+ \mathbf{P}\times \mathbf{K}) .
\label{sr:5}
\eeq
Once the eigenvalues of $M^2$ and $S^2$ are fixed, the representation of
$U(\Lambda,a)$ on the fixed $M^2$ and $S^2$ subspaces of the Hilbert space
is determined by group theoretical considerations.

For this section, it is useful to define the following functions of the
generators:
The Newton-Wigner position operator \cite{Newton:1949cq} is 
\beq
\mathbf{X} := \frac{1}{2} \{ \frac{1}{H},\mathbf{K} \}
- \frac{\mathbf{P} \times (H \mathbf{J} + \mathbf{P} \times \mathbf{K})}{M H (M+H)},
\label{sr:6}
\eeq
and the spin is
\beq
\mathbf{S} :=\mathbf{J} - \mathbf{X}\times \mathbf{P}.
\label{sr:7}
\eeq
These operators satisfy
\beq
[X^i,P^j] = i \delta_{ij} \qquad [X^i,S^j]=[P^i,S^j]=0.
\label{sr:8}
\eeq
Equations (\ref{sr:4}), (\ref{sr:6}) and (\ref{sr:7}) can be inverted to express the ten Poincar\'e
generators as functions of
$\{ M^2,\mathbf{P},\mathbf{X},\mathbf{S}\} $:
\beq
H = \sqrt{\mathbf{P}^2+M^2}
\label{sr:9}
\eeq
\beq
\mathbf{J}= \mathbf{X}\times \mathbf{P} + \mathbf{S}
\label{sr:10}
\eeq
\beq
\mathbf{K} = \frac{1}{2}\{H,\mathbf{X}\} - \frac{\mathbf{P}\times\mathbf{S}}{H+M}.
\label{sr:11}
\eeq
The Poincar\'e commutation relations follow from
(\ref{sr:8}) and the requirement that $M$ and $\mathbf{S}^2$ commute with these operators. 

Bound states of the $N$-particle system are simultaneous eigenstates of
the mass $M$, linear momentum $\mathbf{P}$, the square of the spin
$\mathbf{S}^2$, and the projection of the spin on an axis,
$\mathbf{S}\cdot \hat{\mathbf{z}}$, where the mass eigenvalue $m_b$ is discrete 
.  These states, denoted by
\beq
\vert (m_b,s) \, \mathbf{p},\mu \rangle ,
\label{sr:12}
\eeq
are eigenstates of $H$ with eigenvalue
\beq
E_b = \sqrt{\mathbf{p}^2 + m_b^2}.   
\label{sr:13}
\eeq
Poincar\'e transformations on these states leave $m_b$ and $s$ unchanged:
\[
U(\Lambda,a) \vert (m_b,s) \, \mathbf{p},\mu \rangle 
\]
\beq
=\sum_{\mu'}\int  \vert (m_b,s) \, \mathbf{p}',\mu' \rangle \, d\mathbf{p}' \,
\langle (m_b,s) \, \mathbf{p}',\mu' \vert U(\Lambda,a)\vert 
(m_b,s) \, \mathbf{p},\mu \rangle ,
\label{sr:13a}
\eeq
where the matrix
\[
\langle (m_b,s) \, \mathbf{p}',\mu' \vert U(\Lambda,a)\vert 
(m_b,s) \, \mathbf{p},\mu \rangle
\]
is a representation of a mass $m_b$ and spin $s$ irreducible representation
of the Poincar\'e group. It is the Poincar\'e group analog \cite{Keister:1991sb} of the
Wigner $D$-function for the rotation group:
\[
D^s_{\mu\mu'}(R):=  \langle s, \mu \vert U(R) \vert s, \mu' \rangle .
\]

In a relativistic quantum theory, the cluster condition
\beq
\lim_{\vert \mathbf{x}_i -\mathbf{x}_j \vert \to \infty} T^{\dagger}(\mathbf{x}_1, \cdots, \mathbf{x}_{n_a}) \, U(t) \,
T(\mathbf{x}_1, \cdots, \mathbf{x}_{n_a})
= \otimes_{i=1}^{n_a} U_{a_i}(t)
\label{sr:14}
\eeq
is replaced by 
\beq
\lim_{\vert \mathbf{x}_i -\mathbf{x}_j \vert \to \infty} T^{\dagger}(\mathbf{x}_1, \cdots, \mathbf{x}_{n_b}) \, U(\Lambda ,a) \,
T(\mathbf{x}_1, \cdots, \mathbf{x}_{n_b})
= \otimes_{i=1}^{n_b} U_{b_i}(\Lambda ,a)  = U_{b}(\Lambda ,a) ,
\label{sr:15}
\eeq
where $U_{b_i}(\Lambda ,a)$ are unitary representations of the Poincar\'e group
for the subsystem of particles in the $i^{th}$ cluster of $b$ and the limit is a
strong limit. This condition
means that
it is possible to test special relativity on isolated subsystems of particles.

The new complication in the relativistic case is that interactions
necessarily appear in more than one of the generators \cite{Dirac:1949cp}.
This is a consequence of the commutators
\beq
[K^i,P^i] = i\delta^{ij} H
\label{sr:17}
\eeq
that have the Hamiltonian on the right side. If $H$ includes
interaction terms, then they must also appear in the terms on the left-hand side of
these commutators.
This impacts (\ref{sr:15}) since the translation generators
$\mathbf{P}_{a_i}$ for the $i^{th}$ cluster of $a$ do not commute
with the corresponding boost generator $\mathbf{K}_{a_i}$.

The cluster condition in (\ref{sr:15}) will be satisfied for short-range
interactions provided that each Poincar\'e generator $G_i$ has a
cluster expansion of the form
\beq
G_i = \sum_{a \in {\cal P}_{N}} [G_i]_a = [G_i]_1 + \sum_{a \in {\cal P}_{N}'}{\cal C}_a (G_i)_a ,  
\label{sr:16}
\eeq
where $(G_i)_{a} = \sum_{l=1} ^{n_a} (G_i)_{a_l}$. Each $(G_i)_{a_l}$ has the form
$(\tilde{G}_i)_{a_l}\otimes I $, where  $(\tilde{G}_i)_{a_l}$ only acts on the
Hilbert space associated with the particles in the $l^{th}$ cluster of $a$
and satisfies
the Poincar\'e commutation relations for each cluster of the partition $a$.
$I$ is the identity on the rest of the Hilbert space.
The construction of unitary representations of the Poincar\'e group
consistent with the cluster condition in (\ref{sr:15}) is non-trivial and
can be found in
\cite{Sokolov:1977}\cite{Coester:1982vt}.

The construction in \cite{Coester:1982vt} is recursive on the number of particles.
It uses sums of proper subsystem generators to construct
the Poincar\'e generators 
$(G_i)_a$.  These are used in equations (\ref{sr:4}), (\ref{sr:6}) and (\ref{sr:7})
to construct the operators $M_a,\mathbf{P}_a, \mathbf{X}_a$, and
$\mathbf{S}_a$. For each partition $a$, a $S$-matrix preserving
unitary transformation, $V(a)$, is constructed that transforms
\beq
\mathbf{P}_a, \mathbf{X}_a, \mbox{ and } \mathbf{S}_a
\label{sr:18}
\eeq
to
\beq
\mathbf{P}_0, \mathbf{X}_0, \mbox{ and } \mathbf{S}_0 .
\label{sr:19}
\eeq
The operators with $0$ subscripts have no interactions,
and $V(a)$ is recursively constructed to satisfy
\beq
(V(a))_b = V(a\cap b).
\label{sr:20}
\eeq
The resulting transformed mass operators
\beq
V^{\dagger}(a) \, M_a \, V(a),
\eeq
for each partition,  commute with the operators in (\ref{sr:19}).
If these are combined using
\beq
\tilde{M}= \sum_{a \in {\cal P}_{N}'} {\cal C}_a V^{\dagger}(a) \, M_a \, V(a) ,
\label{sr:21}
\eeq
then $\tilde{M}$ also commutes with the operators in (\ref{sr:19}).
The generators $\tilde{G}_i$ can be constructed as functions of
\beq
\tilde{M}, \mathbf{P}_0, \mathbf{X}_0, \mbox {and } \mathbf{S}_0
\label{sr:22}
\eeq
using relations (\ref{sr:9})-(\ref{sr:11}) to express the generators
in terms of the operators in (\ref{sr:22}).  This gives a dynamical
representation of the Poincar\'e Lie algebra. The problem is that
if the interactions between particles in different clusters of $a$ are turned off, then
\[
\tilde{G}_i \to V^{\dagger}(a) \, ({G}_i)_a \, V(a) .
\]
This means that it will violate cluster properties.  The violation of
cluster properties typically involves interactions disappearing that
should not disappear when subsystems are separated.

In order to restore cluster properties without breaking the commutation relations, one
defines the Cayley transform, $K(a)$,
of $V(a)$ by
\beq
K(a) = i (V(a)-I)(I+V(a))^{-1} \qquad \mbox{with} \qquad V(a) = (I -i K(a) )(I+i K(a))^{-1} .   
\label{sr:23}
\eeq
One can define the unitary operator $V$ in terms of these Cayley transforms by
\beq  
V := (I -i \sum_{a \in {\cal P}_{N}'} {\cal C}_a K(a) )(I+i \sum_{a \in {\cal P}_{N}'} {\cal C}_aK(a))^{-1} .
\label{sr:24}
\eeq
This operator has the property that when the interactions between the clusters of $a$ are turned off, it becomes $V(a)$:
\beq
V_a = V(a).
\eeq
Since $V$ is unitary, it follows that
\beq
G_i := V^{\dagger} \tilde{G}_i V
\eeq
satisfies the Poincar\'e commutation relations and 
satisfies
the cluster properties in (\ref{sr:15})-(\ref{sr:16}).
It also follows that  
\beq
U(\Lambda, a) := V^{\dagger} \tilde{U}(\Lambda, a) V 
\label{sr:25}
\eeq
satisfies the cluster properties
\beq
\lim_{\vert \mathbf{x}_i -\mathbf{x}_j \vert \to \infty} T^{\dagger}(\mathbf{x}_1, \cdots, \mathbf{x}_{n_a}) \, U(\Lambda ,a) \, 
T(\mathbf{x}_1, \cdots, \mathbf{x}_{n_a})
= \otimes_{i=1}^{n_a} U_{a_i}(\Lambda ,a).
\label{sr:26}
\eeq
One consequence of this construction is that 
\beq
G_{i} = [G_i]_1 + \sum_{a \in {\cal P}_{N}'} {\cal C}_a (G_i)_a ,
\label{sr:27}
\eeq
where the connected term is generated by the operator
$V$ and is needed to restore the commutation relations.
 
The construction of the operators $V(a)$
in \cite{Coester:1982vt} 
uses the same methods
discussed in section \ref{sec:2}.
The construction outlined above can be performed
by replacing
\beq
\{\tilde{M}, \mathbf{P}_0, \mathbf{X}_0, \mathbf{S}_0\}
\label{sr:28}
\eeq
by
\beq
\{\tilde{M}_I= V^{\dagger}\tilde{M}V, \mathbf{P}_0, \mathbf{X}_I=
V^{\dagger}\mathbf{X}_0 V,
\mathbf{S}_I=
V^{\dagger}\mathbf{S}_0 V \}.
\label{sr:29}
\eeq

As in the non-relativistic case, the starting point is the
construction of channels. Given a partition $a$, there is a scattering channel
$\alpha$ if there are bound states in each cluster of the partition $a$.
In the relativistic case, the states 
in (\ref{sc:5}) 
are replaced by
\beq
\otimes_{i=1}^{n_a}  \vert (m_{b_i}, s_i) \, \mathbf{p}_i,
\mu_i\rangle ,
\label{sc:30}
\eeq
where the $m_{b_i}$ are discrete mass eigenvalues of the
bound state in the $i^{th}$ cluster of $a$.  Note that
these states transform like (\ref{sr:13a}).

Given $U(\Lambda,a)$ satisfying cluster properties, the construction
in the relativistic case is identical to the construction in the
non-relativistic case. The exact projection operator
on the ${\cal A}_1$ scattering channels is 
\beq
P_{{\cal A}_1} =
\sum_{\alpha \in {\cal A}_1}
\Omega^{(-)}(a)\Phi_\alpha \Phi_\alpha^{\dagger} \Omega^{(-) \dagger}(a) ,
\label{sc:31}
\eeq
where the wave operators are the same functions of the Hamiltonian as
in the non-relativistic case.  Cluster properties of $U(\Lambda,a)$
imply that the Hamiltonian and all of the generators have cluster
expansions of the type discussed in section III.  The wave operators
satisfy the general intertwining relations
\[
U(\Lambda ,b) \, \Omega^{(-)}(a) = \Omega^{(-)}(a) \, U_a(\Lambda,b) .  
\]
The projection of the exact generators on the
channel subspace are
\beq
P_{{i \cal A}_1} G_i .
\label{sc:32}
\eeq
These projected Poincar\'e generators  satisfy the Poincar\'e commutation relations because the channel 
projection operators are Poincar\'e invariant. The projected
generators 
have cluster expansions of the general form
\beq
P_{{i\cal A}_1} G_i = [P_{{i\cal A}_1} G_i]_1 +
\sum_{a \in {\cal P}_{N}'} {\cal C}_a (G_{i{\cal A}_1})_a .
\label{sc:33}
\eeq
The term $\sum_{a \in {\cal P}_{N}'} {\cal C}_a (G_{i{\cal A}_1})_a$ can be constructed
from all proper subsystem problems.  While it satisfies cluster properties,
it does not satisfy the commutation relations.
The operators $V(a)$ and $V$ in (\ref{sr:25}) are also constructed
from proper solutions of proper subsystems.

While the $(G_{i{\cal A}_1})_a$ for $a \in {\cal P}_{N}'$  can be constructed
from solutions of proper subsystem problems, without 
the connected term these operators will not satisfy the
Poincar\'e commutation relations.
Connected operators $[G_{i{\cal A}_1}]_1$ that restore
the commutation relations can be constructed directly from the
$(G_{i{\cal A}_1})_a$ for $a \in {\cal P}_{N}'$.
The construction is the same as the one used in the
exact case except the subsystem mass operators are replaced by
the channel projected subsystem mass operators.

Formally,
\[
G_{i{\cal A}_1} =
G_i
\]
\beq
:= V^{\dagger}(\sum_{a \in {\cal P}_{N}'} {\cal C}_a (V(a) \, G_{i{\cal \, A}_1}) V^{\dagger}(a))_a V
=
[G_{{i\cal A}_1}]_1 + \sum_{a \in {\cal P}_{N}'} {\cal C}_a (G_{i{\cal A}_1})_a ,
\eeq
where the connected term is generated by the construction.
The resulting selected channel generators
satisfy the Poincar\'e commutations relations.
The operator $[G_{{i\cal A}_1}]_1$ is not equal to the operator
$[P_{{i\cal A}_1} G_i]_1$ in (\ref{sc:33}), which requires the full
solution of the problem.  The proof of the optical theorem is
identical to the proof in the non-relativistic case.

These $G_{i{\cal A}_1}$ generators can be used to construct the
corresponding unitary representation of the Poincar\'e group.  In this
construction the required part of the $N$-body interaction is frame
dependent.  Note that while the connected term is required, it is not
unique.  Different constructions of the operators $V(a)$ can result in
different $[G_{i{\cal A}_1}]_1$. This is related to the fact that
cluster properties only fix the dynamics up to a $N$-body interaction.

\section{Conclusions}
\label{sec:10}

The role of individual scattering channels in nuclear reactions and
nuclear structure is identified.  The nuclear Hamiltonian is expressed
as a sum over contributions from each scattering channel,  the
individual contributions determine the structure of the exact
scattering wave functions in that channel up to, but not including
$N$-body correlations.  Limiting the sum to a selected subset of channels
results in a truncated Hamiltonian that satisfies an optical theorem
in the selected subset of channels.  No flux is lost to the discarded
channels.  Each time a channel is included in the sum that channel
opens up, and it contributes to the total cross section.

Each scattering channel Hamiltonian is energy independent and impacts
all of the other channels, even at energies where that channel is
closed.  The scattering channel Hamiltonians also determine both the
bound-state spectrum and structure of the bound-state wave functions.

The scattering channel Hamiltonians are constructed by first expressing the
Hamiltonian as sum of tensor products of proper subsystem
Hamiltonians.  This is done using the M\"obius and Zeta functions of the
lattice of partitions that
relate the cluster expansion of the Hamiltonian to a linear
combination of proper subsystem Hamiltonians.

Terms in the spectral decompositions of tensor products of proper
subsystem Hamiltonians are identified with scattering channels.  The
channel Hamiltonians are obtained by changing the order of the sum of
tensor products of proper subsystem Hamiltonians with the sum over
channels.  The result is the part of the full spectral expansion of
the Hamiltonian that can be constructed out of proper subsystem
solutions.  This is the most detailed information about the $N$-body
Hamiltonian that can be obtained without solving the full $N$-body
problem.  Since the terms in the sum involve only proper subsystems,
the channel sum only includes scattering channels and since the channel
expansion is not the full spectral decomposition, the channel
Hamiltonians for different scattering channels do not commute.

Diagonalizing this Hamiltonian generates the $N$-body correlations in
each scattering wave function, generates the bound state wave
functions and determines bound state energies.  Even though all of the
scattering channels are closed at the $N$-body binding energy, in this
representation of the Hamiltonian the bound state binding energies and
wave functions are completely determined by the scattering channel
contributions.  If the original Hamiltonian has no explicit $N$-body
interactions, then $N$-body scattering correlations generated by
diagonalizing the Hamiltonian exactly cancel the bound state
projections.

The representation of the Hamiltonian as a sum of channel Hamiltonians
was originally derived in
(\cite{Polyzou:1978wp}).  There the intended application was the
construction of few-body models of nuclear reactions dominated by a
given set of channels, where flux is conserved and only appears in the
given set of channels.  In this work, time-dependent scattering theory
is used to determine the contribution of different scattering
channels to the full spectral expansion and the structure of bound
states.

In the scattering channel representation, the Hamiltonian is a sum of
many-body operators, even for Hamiltonians with only two-body
interactions.  The integral equations for the scattering states
derived in section \ref{sec:5} are compatible with many-body operators
and can be used to solve for the transition matrix elements for any
selected channel truncation of the Hamiltonian.  When the retained channels
include all channels related by particle exchanges the number of coupled
integral equations is reduced.

The channel representation is developed by applying cluster
expansions to the spectral representation of the exact Hamiltonian.
The construction utilizes results from time-dependent scattering
theory.  The chain rule for wave operators in
\cite{Kato}\cite{bencze:1976} relates the exact solutions to
scattering solutions for Hamiltonians with interactions between
different clusters solutions turned off.  Hunziker's treatment of
cluster properties in scattering \cite{hunziker} provides a framework
for relating the exact and subsystem channel states without having to
use cluster properties of unbounded operators. The approach in this
work has the advantage that the optical theorem in the selected (or
remaining) channels can be understood from the solved form of the
optical theorem for the exact channel projected Hamiltonian.

In this framework, there are no restrictions on
the choice of contributing channels.  The key properties that are
special about this decomposition include: 1) both the retained and
discarded channel Hamiltonians satisfy optical theorems in
complementary sets of channels, 2) the scattering wave functions for
the retained set and discarded set of channels differ from the exact
scattering wave functions only by $N$-body correlations and 3) the
channel truncated Hamiltonians are energy independent.  Because the
result of this decomposition is a Hamiltonian, it is compatible with any
computational method. And because the exact Hamiltonian can be
expressed as a sum of parts with complementary sets of channels, it
provides a framework for investigating the dynamics due to the
excluded channels.  It is also compatible with Hamiltonians that have
many-body interactions that come from effective field theory
\cite{Epelbaum:2002vt}\cite{Epelbaum:1999dj} or unitary scattering
equivalences
\cite{Ekstein:1960}\cite{Glockle:1990}\cite{Bogner:2006pc}\cite{polyzou_equiv}.
The channel decomposition can also be applied to model bound systems
when the reactions are dominated by few-body channels.

The general method can also be applied to relativistic models.  The
application requires Poincar\'e generators that satisfy cluster
properties. While this is a non-trivial constraint
\cite{Sokolov:1977}\cite{Coester:1982vt}, once it is satisfied the
construction proceeds in the same way as in the non-relativistic
construction; however an additional $N$-body operator, constructed from the truncated generators, is needed to restore the Poincar\'e commutation relations.
The relativistic treatment is applicable to models with
color confinement, where color-singlets naturally cluster and the
constituent particles have relativistic energies.  In this case, the
internal structure of the elementary color
singlets is required input.

While the structure of channel truncated Hamiltonians involves
contributions from many subsystems with different signs, the
combinatorial factors
\cite{rota:1995}\cite{Polyzou:1979wf}\cite{Kowalski:1980cg} ensure
that there is no over counting and that the continuous spectrum is
bounded from below.
This is relevant for treating
overlapping channels and systems with identical particles where it is
necessary to retain all channels related to the selected channels
${\cal A}_1$ by exchange of identical particles. This needs to be
done in a manner that avoids over counting.

One of the issues with approximations that preserve the optical
theorem by discarding open channels is that the exact incoming and
outgoing scattering states can live on different subspaces of the
Hilbert space.  The incoming and outgoing states are related by
time-reversal, which implies that approximations that preserve the
optical theorem by discarding open channels may violate time-reversal
invariance.  In scattering applications, the scattering matrix element
involves limits with the general structure
\beq S_{\beta\alpha}=
\lim_{t \to \infty} \Phi_{\beta}^{\dagger} \, e^{iH_b t} e^{-2iHt}
e^{iH_at} \, \Phi_\alpha ,
\eeq 
where the time evolution in $H$ is in the forward direction. So for
the purpose of this work, there is a preferred direction in time. Note
that even though $H_{{\cal A}_1}$ may not be time-reversal invariant,
it is still a Hermitian operator with an absolutely continuous
spectrum that is bounded from below.

\bibliographystyle{spphys}
\bibliography{Reaction_Theory_References.bib}
\end{document}